\documentclass[12pt]{article}
\usepackage{latexsym}
\usepackage{amsmath,amssymb}
\usepackage{bm}
\usepackage{graphicx,xcolor}
\usepackage{amssymb}
\usepackage{cite}
\usepackage{amsfonts, amscd, amsthm}
\usepackage[all]{xy}
\usepackage{here}
\usepackage{ulem}
\usepackage{mathabx}
\usepackage[titletoc,title]{appendix}
\usepackage{mathbbol}

\renewcommand{\theequation}{\thesection.\arabic{equation}}

\newcommand{\dd}{{\rm d}}

\newcommand{\R}{\mathbb{R}}
\newcommand{\Z}{\mathbb{Z}}
\newcommand{\wwp}{\widehat{\wp}}

\makeatletter
\@addtoreset{equation}{section}
\renewcommand{\theequation}{\thesection.\@arabic\c@equation}
\makeatother

\makeatletter
\renewcommand\appendix{\par%\newpage
  \setcounter{section}{0}%
  \setcounter{subsection}{0}%
  \gdef\thesection{Appendix \@Alph\c@section }
  \renewcommand{\theequation}
  {\Alph{section}.\arabic{equation}}
}
\makeatother

\makeatletter
\newcounter{subeqncnt}
\def\thesubeqncnt{\alph{subeqncnt}}
\def\subequations{\begingroup%
\stepcounter{equation}\edef\@tempa{\theequation}%
\let\c@equation\c@subeqncnt\c@subeqncnt\z@
\edef\theequation{\@tempa\noexpand\thesubeqncnt}}

\makeatother
\allowdisplaybreaks

\begin{document}

\titlepage

%%%%%%%%%%%%%%%%%%%%%%%%%%%%%%%%%%%%%%%%%%%%%%%%%%%%%%%%%%%%%%%%%%%%
\title{The Lie Group Structure\\ of Elliptic/Hyperelliptic $\wp$ Functions
} 
%%%%%%%%%%%%%%%
\author{Masahito Hayashi\thanks{masahito.hayashi@oit.ac.jp}\\
Osaka Institute of Technology, Osaka 535-8585, Japan\\
Kazuyasu Shigemoto\thanks{shigemot@tezukayama-u.ac.jp} \\
Tezukayama University, Nara 631-8501, Japan\\
Takuya Tsukioka\thanks{tsukioka@bukkyo-u.ac.jp}\\
Bukkyo University, Kyoto 603-8301, Japan\\
}
\date{\empty}

%%%%%%%%%%%%%%%%%%%%%%%%%%%%%%%%%%%%%%%%%%%%%%%%%%%%%%%%%

\maketitle
\abstract{%
We consider the generalized dual transformation for elliptic/hyperelliptic $\wp$ functions up to genus three. For the genus one case, from the algebraic addition formula, we deduce that the Weierstrass $\wp$ function has the SO(2,1) $\cong$ Sp(2,$\R$)/$\Z_2$ Lie group structure. For the genus two case, by constructing a quadratic invariant form, we find that hyperelliptic $\wp$ functions have the SO(3,2) $\cong$ Sp(4,$\R$)/$\Z_2$ Lie group structure. Making use of quadratic invariant forms reveals that hyperelliptic $\wp$ functions with genus three have the SO(9,6) Lie group and/or it's subgroup structure. 
}

%\vspace{10mm}
%%%%%%%%%%%%%%%%%%%%%%%%%%%%%%%%%%%%%%%%%%%%%%%%%%%%%%%%%%%%%%%%%%%%%%
%%%%%%%%%%%%%%%%%%%%%%%%%%%%%%%%%%%%%%%%%%%%%%%%%%%%%%%%%%%%%%%%%%%%%%
\section{Introduction} 
\setcounter{equation}{0}

Some special type of non-linear differential equations can be solved exactly and further provide a series of infinitely many solutions. We are interested in those ``solvable/integrable'' mechanisms. 

Soliton equations are examples of such equations, hence various methods for studying soliton systems are beneficial for our objective. Starting from the inverse scattering method~\cite{Gardner,Lax,Zakhrov},  the soliton theory has many interesting developments,  such as the AKNS formulation~\cite{Ablowitz}, geometrical approach~\cite{Bianchi,Hermann,Sasaki,Reyes}, B\"{a}cklund transformation~\cite{Wahlquist,Wadati1,Wadati2}, Hirota equation~\cite{Hirota1,Hirota2}, Sato theory~\cite{Sato}, vertex construction of the soliton solution~\cite{Miwa1,Date1,Jimbo1}, and Schwarzian type mKdV/KdV equation~\cite{Weiss}. 

Non-linear integrable models imply the existence of the potential. KdV equation which is a typical soliton equation has a solution of the Weierstrass $\wp$ function. The $\tau$ function is considered as a potential of the KdV equation in the form $u(x-vt)=-2 \partial_x^2 \log \tau(x-vt)$, and the $\tau$ function corresponds to the $\sigma$ function of the Weierstrass $\wp$ function, $\wp(u)=-\partial_u^2 \log \sigma(u)$. Thus, differential equations of hyperelliptic $\wp$ functions are the natural generalization of higher dimensional non-linear integrable models, and the $\sigma$ function plays a role of potential. Hence,  hyperelliptic $\wp$ functions are expected to have an optimal property to examine Lie group structures. 

We expect that there is a Lie group structure behind some non-linear differential equations, which may be a reason why such non-linear differential equations have infinitely many solutions. Here an addition formula of the Lie group structure might be essential. As the representation of the addition formula of the Lie group, algebraic functions such as trigonometric/elliptic/hyperelliptic functions will emerge for solutions of special differential equations.

The AKNS formalism for the Lax pair is a  powerful tool to examine the Lie algebra structure of soliton equations in non-linear integrable models. In our previous researches, we deduced the SO(2,1) $\cong$ Sp(2,$\mathbb{R}$)/$\Z_2$ Lie algebra structure for two-dimensional KdV/ mKdV/ sinh-Gordon models~\cite{Hayashi1,Hayashi2,Hayashi3,Hayashi4,Hayashi5}. Owing to the fact that the KdV equation has the solution of the elliptic $\wp$ function, we deduced that the genus one elliptic $\wp$ function had the SO(2,1) $\cong$ Sp(2,$\mathbb{R}$)/$\Z_2$ Lie algebra structure. In addition, observing the SO(3,2) $\cong$ Sp(4,$\mathbb{R}$)/$\Z_2$ Lie algebra structure for the two-flows (two-dimensional) Kowalevski top~\cite{Hayashi7}, we found that genus two hyperelliptic $\wp$ functions possessed the SO(3,2) $\cong$ Sp(4,$\mathbb{R}$)/$\Z_2$ Lie algebra structure. By directly using the algebraic addition formula of genus two $\wp$ functions, we obtained the degree two Sp(4,$\mathbb{R}$) Lie group structure~\cite{Hayashi8}.

For the general hyperelliptic differential equations, the Lax pair, especially the AKNS formalism, is not known. Thus we directly study the algebraic addition formula and differential equations themselves to find Lie group structure behind. In this study, we use the generalized dual transformation (GDT) to study Lie group structures of elliptic/hyperelliptic $\wp$ functions. The paper is organized as follows: In section $2$, we study the genus one case of elliptic/hyperelliptic $\wp$ functions. We show the differential equation of the Weierstrass $\wp$ function becomes invariant under GDT. By the algebraic addition formula, we deduce that the Weierstrass $\wp$ function has the SO(2,1) $\cong$ Sp(2,$\R$)/$\Z_2$ Lie group structure. In section $3$, we study the genus two case. Differential equations of genus two hyperelliptic $\wp$ functions transform covariantly under GDT. By constructing quadratic invariant forms under GDT, we deduce that hyperelliptic $\wp$ functions have the SO(3,2) $\cong$ Sp(4,$\R$)/$\Z_2$ Lie group structure. In section $4$, we study the genus three case. Differential equations of genus three hyperelliptic $\wp$ functions also transform covariantly under GDT. Making use of quadratic invariant forms under GDT, we prove that hyperelliptic $\wp$ functions have the SO(9,6) Lie group and/or it's subgroup structure. We devote the final section to the summary and discussions.

%\newpage
%\vspace{10mm}
%%%%%%%%%%%%%%%%%%%%%%%%%%%%%%%%%%%%%%%%%%%%%%%%%%%%%%%%%%%%%%%%%%%%%%%%%%
%%%%%%%%%%%%%%%%%%%%%%%%%%%%%%%%%%%%%%%%%%%%%%%%%%%%%%%%%%%%%%%%%%%%%%%%%%
\section{
The Sp(2,$\mathbb{R}$)/$\Z_2$ $\cong$ SO(2,1) Lie group structure of the Weierstrass $\wp$ function}
\setcounter{equation}{0}
In this section, we consider the elliptic curve on $\R$ and parametrize in the form
\begin{equation}
y^2=\sum_{n=0}^4 \lambda_n x^n=
\sum_{n=0}^4 {}_4 C_n a_n x^n=a_4 x^4+4 a_3 x^3+6 a_2 x^2+4a_1 x+a_0, 
\label{2e1}
\end{equation}
where we denote the binomial coefficients as $_mC_n$ and we put $a_4=0$ in the end. The Jacobi's inversion problem is the problem to obtain $x$ as the function of $u_1$ by using
\begin{equation}
\dd u_1=\frac{\dd x}{y} .
\label{2e2}
\end{equation}
We define $\wp_{11}$ in the form%
%%%%%%
\footnote{We use the non-standard notation, $\wp_{11}$, for the Weierstrass $\wp$ function, because we use the same notation, $\wp_{ij}$, for higher genus $\wp$ functions.}
%%%%%%
%
\begin{equation}
\wp_{11}(u_1)=\frac{\lambda_3}{4} x  ,
\label{2e3}
\end{equation}
which provides the differential equation of the form
\begin{equation}
\wp_{1111}=6\wp_{11}^2+\lambda_2 \wp_{11}+\frac{\lambda_1 \lambda_3}{8}
=6\wp_{11}^2+6 a_2 \wp_{11}+2 a_1 a_3  .
\label{2e4}
\end{equation}
Another differential equation is given by
\begin{equation}
\wp_{111}^2=4\wp_{11}^3+\lambda_2 \wp_{11}^2
+\frac{\lambda_1 \lambda_3}{4} \wp_{11} +\frac{\lambda_0 {\lambda_3}^2}{16}
=4\wp_{11}^3+6 a_2 \wp_{11}^2+4 a_1 a_3 \wp_{11}+a_0 a_3^2   .
\label{2e5}
\end{equation}
While, from the $\sigma$ function, we define $\widehat{\wp}_{11}$ in the form  
\begin{equation}
\widehat{\wp}_{11} =-\frac{\partial^2 \log \sigma(u_1)}{{\partial u_1^2}}   . 
\label{2e6}
\end{equation}
Though $\dd \wp_{11}(u_1)=\dd \widehat{\wp}_{11}(u_1)$ is satisfied, $\wp_{11}(u_1)$ is different from $\widehat{\wp}_{11}(u_1)$ by a constant term. By the dimensional analysis, we obtain  $[\wwp_{11}]=[ 1/u_1^2]=[y^2/x^2]$, $[a_2]=[y^2/x^2]$, and we put
\begin{eqnarray}
\wp_{11}(u_1)=\widehat{\wp}_{11}(u_1)- k_{11} a_2 , \quad  ( k_{11}=\text{const.} ) .
\label{2e7}
\end{eqnarray}
Substituting Eq.(\ref{2e7}) into Eq.(\ref{2e4}), the shifted differential equation becomes in the form
\begin{equation}
\wwp_{1111}
=
6\wwp_{11}^2
+6(1-2 k_{11})a_2 \wwp_{11}
+6k_{11}(k_{11}-1)a_2^2
+2a_1 a_3  .
\label{2e8}
\end{equation}
The shifted another differential equation becomes in the form
\begin{align}
\wwp_{111}^2=&4\wwp_{11}^3+6(1-2k_{11})\wwp_{11}^2+(12 k_{11} ^2 a_2^2-12 k_{11}  a_2^2 +4 a_1 a_3) \wwp_{11}
\nonumber\\
&+(-4 k_{11} ^3 a_2^3+6 k_{11} ^2 a_2^3 -4 k_{11}  a_1 a_2 a_3 +a_0 a_3^2)  .
\label{2e9}
\end{align}

In our previous paper~\cite{Hayashi6}, we considerd the dual transformation of the form
\begin{equation}
x'=\frac{1}{x} , \quad y'=\frac{y}{x^{g+1}}, \quad \lambda'_{n}=\lambda_{2g+2-n}, 
\label{2e10}
\end{equation}
for the genus $g$ hyperelliptic curve on $\mathbb{R}$ in the form
\begin{equation}
y^2=\lambda_{2g+2} x^{2g+2} +\lambda_{2g+1} x^{2g+1}
+\cdots+\lambda_1 x+\lambda_0 ,
\label{2e11}
\end{equation}
where we take  $\lambda_{2g+2}=0$, $\lambda_0=0$ as the standard form for the elliptic curve. In this research, we consider a generalized dual transformation that makes elliptic/hyperelliptic curves invariant.

For the genus one case, we consider the generalized dual transformation (GDT):
\begin{equation}
x'=\frac{a x-c}{-b x +d} , \quad y'=\frac{y}{(-b x+ d)^2}, \quad 
\textrm{with}\quad ad-bc=1.
\label{2e12}
\end{equation}
We are able to express $x$ and $y$ by $x'$ and $y'$ as
\begin{equation}
  x=\frac{dx'+c}{bx'+a} , \quad y=\frac{y'}{(b x'+a)^2}.
\label{2e13}
\end{equation}
by using $(-bx+d)=1/(bx'+a)$.

By requiring the elliptic curve to be invariant under the GDT, namely,
\begin{equation}
  y'^2 = a'_4x'^4 + 4a'_3x'^3 + 6a'_2x'^2 + 4a'_1x' + a'_0 ,
\label{2e14}
\end{equation}
the following equation is derived.
\begin{align}
a'_4x'^4 + &4a'_3x'^3 + 6a'_2 x'^2 + 4a'_1x'+ a'_0
\nonumber\\
        =& a_4(dx'+c)^4 + 4 a_3(dx'+c)^3(bx'+a) + 6a_2(dx'+c)^2(bx'+a)^2
\nonumber\\
         &+ 4a_1(dx'+c)(bx'+a)^3+ a_0(bx'+a)^4 .
\label{2e15}
\end{align}
Comparing coefficients of powers of $x'^n,(n=0,1, \cdots, 4)$ in both sides of Eq.(\ref{2e15}), we obtain
\begin{align}
  a'_4 &= a_4d^4 + 4a_3bd^3 + 6a_2b^2d^2 + 4a_1b^3d + a_0b^4,
\label{2e16}\\
  a'_3 &= a_4cd^3 + a_3(ad+3bc)d^2 + 3a_2(ad+bc)bd + a_1(3ad+bc)b^2 + a_0ab^3,
\label{2e17}\\
  a'_2 &= a_4c^2d^2 + 2a_3(ad+bc)cd + a_2(a^2d^2+4abcd+b^2c^2) + 2a_1(ad+bc)ab + a_0a^2b^2,
\label{2e18}\\
  a'_1 &= a_4c^3d + a_3(3ad+bc)c^2 + 3a_2(ad+bc)ac + a_1(ad+3bc)a^2 + a_0a^3b,
\label{2e19}\\
  a'_0 &= a_4c^4 + 4a_3a c^3 + 6a_2a^2c^2 + 4a_1a^3c + a_0a^4  .
\label{2e20}
\end{align}

After the transformation, we put $a_4=0$, $a'_4=0$. For example, we can realize $a'_4=0$ by choosing 
\begin{equation}
a_0=-\frac{4 a_3 d^3+6 a_2 b d^2+4 a_1 b^2 d}{b^3} ,
\label{2e21}
\end{equation}
for the given $a_1$, $a_2$, $a_3$, $a$, $b$, $c$ with $d=(b c+1)/a$. An important point is that we put $a'_4=0$ after fixing parameters $a$, $b$, $c$. According to the transformation, the elliptic curve changes and we take the standard form of the elliptic curve in such a way as $a_0$ is given by Eq.(\ref{2e21}). This transformation of $a_n\,(n=0,1, \cdots, 4)$ generally causes changes of moduli parameters. Under this transformation,  we obtain
\begin{equation}
\dd u'_1=\frac{\dd x'}{y'}=\frac{\dd x}{y}=\dd u_1  .
\label{2e22}
\end{equation}
By using relations $\dd u_1=\dd x/y$ and $\dd u'_1=\dd x'/y'$, a functional form of $x$ with respect to $u_1$ is the same as that of $x'$ with respect to $u'_1$. Further, in general, $u'_1$ is different from $u_1$ by a constant term. Hence, we obtain
\begin{align}
&x'=\wp(u'_1, \{a'_i \}),\quad  x=\wp(u_1, \{a_i \}), \
\label{2e23}\\
& u_1'=u_1+k, \quad (k=\text{const.}) ,
\label{2e24}
\end{align}
with $\{ a_i\}=\{a_0, a_1,a_2.a_3\}$. Because $\widehat{\wp}_{11}$ is defined from the invariant $\sigma$ function, $\sigma'(u'_1)=\sigma(u_1)$, we obtain
\begin{equation}
\widehat{\wp}'_{11} =-\frac{\partial^2 \log \sigma'(u'_1)}{{\partial {u'_1}^2}}
=-\frac{\partial^2 \log \sigma(u_1)}{{\partial {u'_1}^2}}
=\left(\frac{\partial u_1}{\partial u'_1}\right)^2\widehat{\wp}_{11}=\widehat{\wp}_{11} ,
\label{2e25}
\end{equation}
which shows the invariance of $\wwp_{11}$ under the transformation. We must notice that the transformation of $\widehat{\wp}_{11}$ and $a_2$ in Eq.(\ref{2e7}) is different. We impose the differential equation Eq.(\ref{2e8}) to be invariant under the transformation. Sending the coefficient of $\wwp_{11}$ in Eq.(\ref{2e8}) to be zero, we have  $k_{11}=1/2$, so that we obtain the constant shift in the form $\wp_{11}(u_1)=\widehat{\wp}_{11}(u_1)- a_2/2$. Thus, the constant shifted differential equation turns to be 
\begin{equation}
\wwp_{1111}=6\wwp_{11}^2+2 a_1 a_3-\frac{3}{2} a_2^2.
\label{2e26}
\end{equation}
Further, we can prove that $2 a_1 a_3-3 a_2^2/2$ is invariant under the transformation, so that the differential equation is invariant under the transformation. Another constant shifted differential equation is given by
\begin{equation}
\wwp_{111}^2=4\wwp_{11}^3-g_2 \wwp_{11} -g_3, 
\label{2e27}
\end{equation}
with $g_2=3 a_2^2-4 a_1 a_3$ and $g_3=2 a_1 a_2 a_3 -a_2^3-a_0 a_3^2$. Since $g_2$ and $g_3$ are invariant under the transformation, 
we find that another differential equation Eq.(\ref{2e27}) is invariant under the transformation. 

Next, we consider the algebraic addition formula. Using Eq.(\ref{2e23}), Eq.(\ref{2e24}) and the invariance of $g_2$, $g_3$, we obtain
\begin{equation}
 \wwp_{11}(u_1' ,\{g'_2, g'_3\})-\frac{a'_2}{2} 
=\wwp_{11}(u_1+k,\{g _2, g _3\})-\frac{a'_2}{2} 
=\frac{ a\big(\wwp_{11}(u_1, \{g_2, g_3\})-a_2/2\big)-c}
      {-b\big(\wwp_{11}(u_1, \{g_2, g_3\})-a_2/2\big)+d}
\label{2e28}
\end{equation}
We determine $k$ from $\wwp_{11}(k, \{g_2,g_3\})=a'_2/2-a/b$ because of $\wwp_{11}(u, \{g_2, g_3\}) \rightarrow \infty$ as $u \rightarrow 0$. If we adopt Eq.(\ref{2e21}), for the given $a_1$, $a_2$, $a_3$, $a$, $b$, $c$ $(d=(1+b c)/a)$, we can obtain $g_2$, $g_3$, $a'_2$, $k$. Thus, we obtain the left-hand side of Eq.(\ref{2e28}), which is equal to the right-hand side of Eq.(\ref{2e28}) in the SL(2,$\R$) form. This form of the algebraic addition formula of the Weierstrass $\wp$ functions gives the Sp(2,$\mathbb{R}$)/$\Z_2$ $\cong$ SO(2,1) Lie group structure. In this way, in order to consider Lie group structure, we must not fix but change, the elliptic curve in such a way as it becomes invariant under the transformation. From the beginning, if we fix the elliptic curve i.e.\ $a_n=a'_n\, (n=0,1, \cdots, 4)$, Lie group (continuous group) structure does not exist; yet only some discrete group structure exists.

Next, we consider the following special case
\[
  a=d=0,\quad bc=-1.
\]
In this case, we obtain
\begin{equation}
  a'_3=-a_1/c^2,\quad a'_2=a_2,\quad a'_1=-a_3 c^2,\quad a'_0=a_0=0.
\label{2e29}
\end{equation}
Eq.\eqref{2e29} shows $g_2=3 a_2^2-4 a_1 a_3$ and $g_3=2 a_1 a_2 a_3 -a_2^3$ are invariant under the transformation, which is equivalent to the expression of the $g_2, g_3, a_2$ fixed elliptic curve in Eq.(\ref{2e28}). Hence, we obtain
\begin{equation}
\wwp_{11}(u_1+k, \{g_2, g_3\})-\frac{a_2}{2}
=-\frac{c^2}
{\wwp_{11}(u_1, \{g_2, g_3\})-a_2/2}  , 
\label{2e30}
\end{equation}
which provides 
\begin{equation}
\wwp_{11}(u_1+k, \{g_2, g_3\})=\frac{(a_2/2)\ \wwp_{11}(u_1, \{g_2, g_3\})-a_2^2/4-c^2}
{\wwp_{11}(u_1, \{g_2, g_3\})-a_2/2}  .
\label{2e31}
\end{equation}

Using $e_1$, $e_2$ and $e_3$, we parametrize $g_2$ and $g_3$ in the standard form by solving following equations, 
\begin{align}
 e_1+e_2+e_3 &= 0,
\label{2e32}\\
 e_1e_2 + e_2e_3+e_3e_1 &= -\frac{3}{4}a_2^2+a_1a_3 \left(=-\frac{g_2}{4}\right),
\label{2e33}\\
 e_1e_2 e_3 &= \frac12a_1a_2a_3 -\frac{1}{4}a_2^3 \left(= \frac{g_3}{4}\right).
\label{2e34}
\end{align}
From Eqs.\eqref{2e33}--\eqref{2e34}, we obtain
\begin{equation}
  (a_2-2 e_1)(a_2-2 e_2)(a_2-2 e_3)=0,
\label{2e35}
\end{equation}
and we choose $a_2=2 e_1$. Then $e_2$ and $e_3$ are determined from the following equations
\begin{align}
  e_2+e_3 &= -e_1 = -\frac{a_2}{2},
\label{2e36}\\
  e_2e_3 &=-\frac{a_2^2}{2} + a_1a_3.
\label{2e37}
\end{align}

By using $\wwp_{11}(0,\{g_2, g_3\})\rightarrow \infty$, $k$ is determined from $\wwp(k, \{g_2, g_3\})=e_1$, that is, $k$ is the first half period, namely $k=\omega_1$. Then Eq.(\ref{2e31}) is the half period addition formula of $\wwp_{11}(u,\{g_2, g_3\})$. Thus we must choose $c$ as a special vales $-c^2=(e_1-e_2)(e_1-e_3)$. Then we finally obtain~\cite{Hancock} 
\begin{equation}
\wwp_{11}(u_1+\omega_1, \{g_2, g_3\})
=
\frac{e_1\wwp_{11}(u_1, \{g_2, g_3\})-e_1^2+(e_1-e_2)(e_1-e_3)}
{\wwp_{11}(u_1, \{g_2, g_3\})-e_1}, 
\label{2e38}
\end{equation}
with $g_2=-4(e_1e_2+e_2e_3+e_3e_1)$ and $g_3=4 e_1 e_2 e_3$. Therefore, in this special case, the algebraic addition formula has the order two Sp(2,$\mathbb{R}$) Lie group structure~\cite{Hayashi8}.

Further, we consider more special case of $c=\pm 1$. In this case, the transformation becomes in the form
\begin{eqnarray}
x'=-\frac{1}{x}, \quad y'=\frac{y}{x^2}, \quad  a'_3=-a_1, \quad a'_2=a_2, \quad a'_1=-a_3.
\label{2e39}
\end{eqnarray}
If we further make the transform of the elliptic curve to be invariant in the form $x''=-x'$ , $y''=y$ and $a''_n=(-1)^{n} a'_n$ , we obtain $x''=1/x, \quad y''=y/x^2, \quad  a''_3=a_1, \quad a''_2=a_2, \quad a''_1=a_3$, which is equivalent to the dual transformation Eq.(\ref{2e10}) in the genus one case.

%\newpage
%\vspace{10mm}
%%%%%%%%%%%%%%%%%%%%%%%%%%%%%%%%%%%%%%%%%%%%%%%%%%%%%%%%%%%%%%%%%%%%%%%%%%
%%%%%%%%%%%%%%%%%%%%%%%%%%%%%%%%%%%%%%%%%%%%%%%%%%%%%%%%%%%%%%%%%%%%%%%%%%
%%%%%%%%%%%%%%%%%%%%%%%%%%%%%%%%%%%%%%%%%%%%%%%%%%%%%%%%%%%%%%%%%%%%%%%%%%
\section{
The Sp(4,$\mathbb{R}$)/$\Z_2$ $\cong$ SO(3,2) Lie group structure
of genus two hyperelliptic $\wp_{ij}$ functions} 
\setcounter{equation}{0}
We parametrize the genus two hyperelliptic curve on $\mathbb{R}$ in the form
\begin{equation}
y^2=\sum_{n=0}^6 \lambda_n x^n=\sum_{n=0}^6 {}_6 C_n a_n x^n
=a_6 x^6+6 a_5 x^5+15 a_4 x^4+20 a_3 x^3+15 a_2 x^2+6a_1 x+a_0   ,
\label{3e1}
\end{equation}
where we put $a_6=0$ in the end. The Jacobi's inversion problem is the problem to express the symmetric combination of 
$x_1$ and $x_2$ as the function of $u_1$ and $u_2$ by using relations
\begin{eqnarray}
\dd u_1 =\frac{\dd x_1}{y_1}+\frac{\dd x_2}{y_2}, \quad 
\dd u_2 =\frac{ x_1 \dd x_1}{y_1}+\frac{x_2 \dd x_2}{y_2}   . 
\label{3e2}
\end{eqnarray}
From above relations, we obtain 
\begin{eqnarray}
\frac{\partial x_1}{\partial u_2}=\frac{y_1}{x_1-x_2}, \quad
\frac{\partial x_2}{\partial u_2}=-\frac{y_2}{x_1-x_2},  \quad
\frac{\partial x_1}{\partial u_1}=-\frac{x_2 y_1}{x_1-x_2}, \quad
\frac{\partial x_2}{\partial u_1}=\frac{x_1 y_2}{x_1-x_2}  .  
\label{3e3}
\end{eqnarray}
Thus, we obtain 
\begin{equation}
\frac{\partial (x_1+x_2)}{\partial u_1}=-\frac{\partial (x_1x_2)}{\partial u_2}  .
\label{3e4}
\end{equation}
As the solution of the Jacobi's inversion problem, we define 
\begin{equation}
\wp_{22}(u_1,u_2)=\frac{\lambda_5}{4}(x_1+x_2), \quad
\wp_{21}(u_1,u_2)=-\frac{\lambda_5}{4} x_1 x_2, 
\label{3e5}
\end{equation}
and Eq.(\ref{3e4}) provides the integrability condition of genus two hyperelliptic $\wp$ functions
$$ \frac{\partial \wp_{22}(u_1,u_2)}{\partial u_1}
=\frac{\partial \wp_{21}(u_1,u_2)}{\partial u_2}  .$$
Furthermore, if we define~\cite{Baker1,Baker2}
\begin{align}
\wp_{11}(u_1.u_2)=&\frac{F(x_1,x_2)-2y_1 y_2}{4(x_1-x_2)^2}, 
\label{3e6}\\
F(x_1,x_2)
=&2 a_6 x_1^3 x_2^3+6 a_5 x_1^2 x_2^2 (x_1+x_2)+30 a_4 x_1^2 x_2^2+20 a_3 x_1 x_2 (x_1+x_2)+30 a_2 x_1 x_2
\nonumber\\
&+6 a_1 (x_1+x_2)+2 a_0 ,
\nonumber
\end{align}
we obtain full integrability conditions
\begin{equation}
\frac{\partial \wp_{22}(u_1,u_2)}{\partial u_1}
=\frac{\partial \wp_{21}(u_1,u_2)}{\partial u_2}, \quad
\frac{\partial \wp_{21}(u_1,u_2)}{\partial u_1}
=\frac{\partial \wp_{11}(u_1,u_2)}{\partial u_2}  .
\label{3e7}
\end{equation}

Next, we define another genus two hyperelliptic $\wwp$ functions  constructed  from the $\sigma$ function in the form
\begin{equation}
\widehat{\wp}_{ij}(u_1,u_2) =-\frac{\partial^2 \log \sigma(u_1,u_2)}{{\partial u_i \partial u_j}} .
\label{3e8}
\end{equation}
Though $\wp_{ij}(u_1,u_2)$ and $\widehat{\wp}_{ij}(u_1,u_2)$ satisfy the same integrability conditions
$$
\partial_i \wp_{jk}(u_1,u_2)=\partial_j \wp_{ik}(u_1,u_2) \quad\mathrm{and}\quad \partial_i \wwp_{jk}(u_1,u_2)=\partial_j \wwp_{ik}(u_1,u_2),
$$
$\wp_{ij}(u_1,u_2)$ and $\wwp_{ij}(u_1,u_2)$ are not equal but differ by a constant. By the dimension analysis, we obtain
$[\wp_{22}]=[1/u_2^2  ]=[y^2/x^4]$,
$[\wp_{21}]=[1/u_1 u_2]=[y^2/x^3]$,
$[\wp_{11}]=[1/u_1^2  ]=[y^2/x^2]$,
$[a_4]=[y^2/x^4]$,
$[a_3]=[y^2/x^3]$,
$[a_2]=[y^2/x^2]$,
so that three pairs $(\wp_{22}, a_4)$, $(\wp_{21}, a_3)$, and $(\wp_{11}, a_2)$ have the same dimensions. Thus, we put
\begin{align}
\wp_{22}(u_1,u_2)&=\widehat{\wp}_{22}(u_1,u_2)- k_{22} a_4  ,
\label{3e9}\\
\wp_{21}(u_1,u_2)&=\widehat{\wp}_{21}(u_1,u_2)- k_{21} a_3  ,
\label{3e10}\\
\wp_{11}(u_1,u_2)&=\widehat{\wp}_{11}(u_1,u_2)- k_{11} a_2  ,
\label{3e11}
\end{align}
where $a_i$ are coefficients of the hyperelliptic curve, and $k_{22}, k_{21}, k_{11}$ are some numerical constants. We determine constants $k_{ij}$ in such a way as whole differential equations transform covariantly
%%%%%%
\footnote{In the general coordinate transformation, a tensor $T_{\mu \nu}$ transforms covariantly in the form
$\displaystyle{T'_{\mu \nu}=\frac{\partial x^{\alpha}}{\partial x'^{\mu}} 
                            \frac{\partial x^{\beta }}{\partial x'^{\nu}} T_{\alpha \beta}}$. 
We use the terminology transform covariantly in the same way. Observing a transformation $\displaystyle{\wwp'_{i j}=\frac{\partial u_p}{\partial u'_i}\frac{\partial u_q}{\partial u'_j} \wwp_{p q}}$ which shows up later, we may say that $\wwp_{ij}$ transform covariantly. 
}.
%%%%%%

Let us start with the following differential equations~\cite{Baker2,Baker3},  
\begin{align}
&1)\ \wp_{2222} -6 \wp_{22}^2+3 \lambda_6 \wp_{11} - \lambda_5\wp_{21} -\lambda_4\wp_{22}  
- \frac{1}{8}\lambda_5 \lambda_3+\frac{1}{2}\lambda_6 \lambda_2=0,
\label{3e12}\\
&2)\ \wp_{2221} -6 \wp_{22} \wp_{21}+ \frac{1}{2} \lambda_5\wp_{11} -\lambda_4\wp_{21}
+\frac{1}{4}\lambda_6 \lambda_1=0 ,
\label{3e13}\\
&3)\ \wp_{2211}-4\wp_{21}^2 -2 \wp_{22} \wp_{11} -\frac{1}{2}\lambda_3\wp_{21}+\frac{1}{2}\lambda_6 \lambda_0=0,
\label{3e14}\\
&4)\ \wp_{2111} -6 \wp_{21} \wp_{11}-\lambda_2\wp_{21}+\frac{1}{2}\lambda_1\wp_{22}+\frac{1}{4}\lambda_5 \lambda_0=0,
\label{3e15}\\
&5)\ \wp_{1111} -6 \wp_{11}^2 -\lambda_2\wp_{11}-\lambda_1\wp_{21}+3 \lambda_0 \wp_{22}
-\frac{1}{8}\lambda_3 \lambda_1+\frac{1}{2}\lambda_4 \lambda_0=0  . 
\label{3e16}
\end{align}
We rewrite the above differential equations with $a_n$ instead of $\lambda_n$. Later we will explain how to determine $k_{ij}$, but we provide here the values of them,  $k_{22}=3/2, k_{21}=1/2, k_{11}=3/2$. Thus, the constant shift of $\wp_{ij}$ are given in the form
\begin{equation}
\wp_{22}=\wwp_{22}-\frac{3}{2} a_4,\quad \wp_{21}=\wwp_{21}-\frac{1}{2} a_3,\quad \wp_{11}=\wwp_{11}-\frac{3}{2} a_2 .
\label{3e17}
\end{equation}
Hence, we obtain constant shifted differential equations ~\cite{Baker2}
\begin{align}
&1)'\ \wwp_{2222} -6 \wwp_{22}^2+3\left( a_6 \wwp_{11} -2 a_5\wwp_{21} +a_4\wwp_{22}\right)  
+3\left(a_6 a_2-4 a_5 a_3 +3 a_4^2 \right)=0,
\label{3e18}\\
&2)'\ \wwp_{2221} -6 \wwp_{22} \wp_{21}+ 3 \left( a_5\wwp_{11} -2 a_4\wwp_{21}+a_3 \wwp_{22} \right)
+\frac{3}{2} \left( a_6 a_1-3 a_5 a_2+2 a_4 a_3 \right)  =0 ,
\label{3e19}\\
&3)'\ \wwp_{2211}-(4\wwp_{21}^2 +2 \wwp_{22} \wwp_{11}) 
+3 \left( a_4 \wwp_{11}-2 a_3   \wwp_{21}+a_2 \wwp_{22} \right)
+\frac{1}{2} \left( a_6 a_0 -9 a_4 a_2 +8 a_3^2 \right)  =0,
\label{3e20}\\
&4)'\ \wwp_{2111} -6 \wwp_{21} \wwp_{11}+3 \left( a_3 \wwp_{11} -2 a_2 \wwp_{21} +a_1 \wwp_{22} \right)
+\frac{3}{2} \left(a_5 a_0 -3 a_4 a_1 +2 a_3 a_2 \right)=0,
\label{3e21}\\
&5)'\ \wwp_{1111} -6 \wwp_{11}^2 +3 \left( a_2 \wwp_{11}-2 a_1\wwp_{21}+a_0 \wwp_{22}\right)
+3 \left(a_4 a_0 -4 a_3 a_1 +3 a_2^2 \right)=0   . 
\label{3e22}
\end{align}
From Eqs.(\ref{3e18})--(\ref{3e22}), we define the first, second, third and fourth term of each equation as the component of vectors of ${\bf P}$, ${\bf Q}$, ${\bf R}$ and ${\bf S}$ in the form  
\begin{align}
{\bold P}&=\left( \begin{array}{@{\,}c@{\,}}
\wwp_{2222} \\
\wwp_{2221} \\
\wwp_{2211} \\
\wwp_{2111} \\
\wwp_{1111}
\end{array} \right ), \  
{\bold Q}=\left( \begin{array}{@{\,}c@{\,}}
-6 \wwp_{22}^2 \\
-6 \wwp_{22}\wwp_{21} \\
-4 \wwp_{21}^2 -2 \wwp_{22}\wwp_{11} \\
-6 \wwp_{21}\wwp_{11} \\
-6 \wwp_{11}^2
\end{array} \right ), \ 
{\bold R}=3\left( \begin{array}{@{\,}c@{\,}}
a_6 \wwp_{11}-2 a_5 \wwp_{21}+a_4\wwp_{22} \\
a_5 \wwp_{11}-2 a_4 \wwp_{21}+a_3\wwp_{22} \\
a_4 \wwp_{11}-2 a_3 \wwp_{21}+a_2\wwp_{22} \\
a_3 \wwp_{11}-2 a_2 \wwp_{21}+a_1\wwp_{22} \\
a_2 \wwp_{11}-2 a_1 \wwp_{21}+a_0\wwp_{22} 
\end{array} \right ), 
\nonumber\\
{\bold S}&=3\left( \begin{array}{@{\,}c@{\,}}
 a_6 a_2-4 a_5 a_3+3 a_4^2\\
(a_6 a_1-3 a_5 a_2+2 a_4 a_3)/2\\
(a_6 a_0-9 a_4 a_2+8 a_3^2)/6\\
(a_5 a_0-3 a_4 a_1+2 a_3 a_2)/2 \\
 a_4 a_0-4 a_3 a_1+3 a_2^2
\end{array} \right )   .
\label{3e23}
\end{align}
Each differential equation is given in the form
\begin{equation}
P_n+Q_n+R_n+S_n=0, \ (n=1, 2, \cdots, 5), 
\label{3e24}
\end{equation}
which provide Eqs.(\ref{3e18})--(\ref{3e22}). 

Next, we consider the generalized dual transformation of the form
\begin{equation}
x'=\frac{a x -c}{-b x +d} , \quad y'=\frac{y}{(-b x+ d)^3}, \quad 
\textrm{with} \quad ad-bc=1, 
\label{3e25}
\end{equation}
in such a way as such transformation makes the hyperelliptic curve on $\mathbb{R}$ to be invariant. Then $a'_n \ (n=1, 2, \cdots, 6)$ are systematically determined from the relation
\begin{equation}
\sum_{n=0}^6 {}_6 C_n a_n (b x' +a)^{6-n} (d x'+c)^{n}=\sum_{n=0}^6 {}_6 C_n a'_n x'^n.
\label{3e26}
\end{equation}
The explicit expressions of the transformations of $a_n$ are given in Appendix A. We put $a_6=0$ and $a'_6=0$ after the transformation. The transformed Jacobi's inversion relations are given by
\begin{align}
\dd u'_1 &=\frac{\dd x'_1}{y'_1}+\frac{\dd x'_2}{y'_2}
=\frac{(-b x_1 +d) \dd x_1}{y_1}+\frac{(-b x_2 +d) \dd x_2}{y_2}
=d\ \dd u_1 -b\ \dd u_2 ,
\label{3e27}\\
\dd u'_2 &=\frac{x'_1 \dd x'_1}{y'_1}+\frac{x'_2 \dd x'_2}{y'_2}
=\frac{(a x_1 -c) \dd x_1}{y_1}+\frac{(a x_2 -c) \dd x_2}{y_2}
=-c\ \dd u_1+a\ \dd u_2. 
\label{3e28}
\end{align}
Then we obtain 
\begin{equation}
\frac{\partial}{\partial u'_1}=a \frac{\partial}{\partial u_1}+c \frac{\partial}{\partial u_2}, \quad
\frac{\partial}{\partial u'_2}=b \frac{\partial}{\partial u_1}+d \frac{\partial}{\partial u_2}  . 
\label{3e29}
\end{equation}
We require that the hyperelliptic curve becomes invariant under the transformation, which implies that the $\sigma$ function is invariant. From the invariance of the $\sigma$ function under the transformation, $\sigma'(u'_1, u'_2)=\sigma(u_1, u_2)$, the transformed $\widehat{\wp}_{ij}$ functions are given by
$$
\displaystyle{\widehat{\wp}'_{i j}(u'_1,u'_2)
=-\frac{\partial^2 \log \sigma'(u_1, u_2)}{\partial u'_i \partial u'_i}
=-\frac{\partial^2 \log \sigma (u_1, u_2)}{\partial u'_i \partial u'_i}}.
$$
Then $\wwp_{ij}$ transform in covariant forms, 
\begin{align}
&\wwp'_{22}=\frac{\partial u_p}{\partial u'_2}\frac{\partial u_q}{\partial u'_2}\wwp_{p q}
=d^2 \wwp_{22}+2 b d \wwp_{21}+b^2 \wwp_{11}  , 
\label{3e30}\\
&\wwp'_{21}=\frac{\partial u_p}{\partial u'_2}\frac{\partial u_q}{\partial u'_1}\wwp_{p q}
=c d \wwp_{22}+(a d+b c)\wwp_{21}+a b \wwp_{11}  ,
\label{3e31}\\
&\wwp'_{11}=\frac{\partial u_p}{\partial u'_1}\frac{\partial u_q}{\partial u'_1}\wwp_{p q}
=c^2 \wwp_{22}+2 a c \wwp_{21}+a^2 \wwp_{11}  .
\label{3e32}
\end{align}
A simple rule to obtain the above result is as follows. From Eq.(\ref{3e29}), we consider $P'_1=a P_1+c P_2$ and $P'_2=b P_1+d P_2$.
Making ${P'_2}^2=b^2 P_1^2+2 b d P_1 P_2+d^2 P_2^2$ and replace 
$$
{P'_2}^2 \rightarrow \wwp'_{22},\quad
 P_2^2   \rightarrow \wwp_{22},\quad
 P_2 P_1 \rightarrow \wwp_{21},\quad
 P_1^2   \rightarrow \wwp_{11},
$$
which gives Eq.(\ref{3e30}). This simplified rule is useful to obtain transformed expressions of $\wwp'_{ijk\ell}$ by considering $P'_i P'_j P'_k P'_{\ell}$. Thus $\wwp_{ijk\ell}$ transform in the covariant form
\begin{equation}
\wwp'_{ijk\ell}=\frac{\partial u_p}{\partial u_i'}\frac{\partial u_q}{\partial u_j'}
\frac{\partial u_r}{\partial u_k'} \frac{\partial u_s}{\partial u_{\ell}'} \wwp_{pqrs},
\label{3e33} 
\end{equation}
which provides
\begin{equation}
\left( \begin{array}{@{}c@{}}
\wwp'_{2222} \\
\wwp'_{2221} \\
\wwp'_{2211} \\
\wwp'_{2111} \\
\wwp'_{1111}
 \end{array} \right )
=\left( \begin{array}{@{}ccccc@{}}
    d^4 &        4b d^3 &       6 b^2d^2     &        4 b^3d&    b^4 \\
 c  d^3 &  ( ad+3bc)d^2 &3( ad+bc)b  d       & (3ad+ bc)b^2 & a  b^3 \\
 c^2d^2 & 2( ad+ bc)cd  & a^2d^2+4abcd+b^2c^2& 2(ad+bc) ab  & a^2b^2 \\
 c^3d   &  (3ad+ bc)c^2 &3( ad+bc)a  c       & ( ad+3bc)a^2 & a^3b   \\
 c^4    &        4a c^3 &       6 a^2c^2     &        4 a^3c& a^4    \\
 \end{array} \right)
\left( \begin{array}{@{}c@{}}
\wwp_{2222} \\
\wwp_{2221} \\
\wwp_{2211} \\
\wwp_{2111} \\
\wwp_{1111}
\end{array} \right ) .
\label{3e34}
\end{equation}
We denote this as ${\bold P}'=M {\bold P}$. Then, by the same $M$, we can prove ${\bold Q}'=M {\bold Q}$. We determined $k_{ij}$ in Eqs.(\ref{3e9})--(\ref{3e11}) in such a way as ${\bold R}$ transform in the same way as ${\bold R}'=M {\bold R}$, so that $k_{22}=3/2$, $k_{21}=1/2$ and $k_{11}=3/2$ can be obtained. Thus, as we have promised, the constant shift of $\wp_{ij}$ has been determined as Eq.(\ref{3e17}). Hence, by using $M$ and $k_{ij}$, we can prove ${\bold S}'=M {\bold S}$. We define the total vector ${\bf T}={\bf P}+{\bf Q}+{\bf R}+{\bf S}$, whole differential equations transform covariantly in the form ${\bold T}'=M {\bold T}$. From differential equations ${\bf T}=0$, we obtain ${\bf T'}=0$, that is, the set of differential equations ${\bf T}=0$ is invariant.

We have shown that differential equations have the Lie group (continuous group) structure; yet an issue is what type of Lie group structure the differential equations have. To elucidate the problem, we try to find quadratic invariances defined from some vector $\bf X$. We here adopt ${\bf X}={\bf P}$, that is,
$$
X_1=\wwp_{2222},\quad
X_2=\wwp_{2221},\quad
X_3=\wwp_{2211},\quad
X_4=\wwp_{2111},\quad
X_5=\wwp_{1111}.
$$
Since the dual transformation is a special case of the generalized transformation, the invariance of the following dual transformations 
$$
\wwp_{2222} \leftrightarrow \wwp_{1111},\quad
\wwp_{2221} \leftrightarrow \wwp_{2111},\quad
\wwp_{2211} \leftrightarrow \wwp_{2211}
$$
are necessary. Thus, as the quadratic invariance, we obtain
\begin{equation}
I=\ell_1 \wwp_{2222} \wwp_{1111}+\ell_2 \wwp_{2221} \wwp_{2111}+\ell_3 \wwp_{2211}^2
=\ell_1 X_1 X_5+\ell_2 X_2 X_4+\ell_3 X_3^2 .
\label{3e35}
\end{equation}
By imposing the invariance under the transformation, the coefficients $\ell_1, \ell_2, \ell_3$ are determined to be 
\begin{equation}
I=\wwp_{2222} \wwp_{1111}-4\wwp_{2221} \wwp_{2111}+3\wwp_{2211}^2
=X_1 X_5-4 X_2 X_4+3 X_3^2=\text{inv.} .
\label{3e36} 
\end{equation}
Even if we adopt ${\bf X}$ as any of $\{ {\bf P},  {\bf Q},  {\bf R},  {\bf S} \}$, Eq.(\ref{3e36}) gives invariants. Thus, we define
\begin{equation}
X_1=Y_1+Y_4,\quad 
X_2=\frac{Y_5+Y_2}{2},\quad 
X_3=\frac{Y_3}{\sqrt{3}},\quad 
X_4=\frac{Y_5-Y_2}{2},\quad 
X_5=Y_1-Y_4,
\label{3e37}
\end{equation}
and we arrive at the quadratic invariance of the form
\begin{equation}
 I=\left(Y_1^2+Y_2^2+Y_3^2\right)-\left(Y_4^2+Y_5^2\right)=\text{inv.}
\label{3e38}
\end{equation}
Therefore, we conclude that differential equations have the SO$(3,2)$ $\cong$Sp(4,$\mathbb{R}$)/$\Z_2$ Lie group structure, which is consistent with our previous results\cite{Hayashi7,Hayashi8}. At this level, we put $a_6=0$ and $a'_6=0$, and $a'_6=0$ is realized by taking the standard form of the hyperelliptic curve with $a_0=-(6 a_5 d^5+15 a_4 b d^4+20 a_3 b^2 d^3 +15 a_2 b^3 d^2+6 a_1 b^4 d)/b^5$. As the invariance of the transformation is identically satisfied, we obtain the same result even if we put constraints $a_6=0$ and $a'_6=0$.

%%%%%%%%%%%%%%%%%%%%%%%%%%%%%%%%%%%%%%%%%%
\subsection{The Lie algebraic approach to the constant shift of $\wp_{ij}$ and the quadratic invariant}

We consider the following three infinitesimal transformations derived from Eq.(\ref{3e25})~\cite{Baker5}, 
\begin{alignat}{2}
{\rm i)}  \ &x'= x + \epsilon,    &\qquad(a&=1,\ b=0,\ c=-\epsilon,\ d=1)
\label{3e39}\\
{\rm ii)} \ &x'= x + \epsilon x,  &      (a&=1+\epsilon/2,\ b=0,\ c=0,\ d=1-\epsilon/2)
\label{3e40}\\
{\rm iii)}\ &x'= x + \epsilon x^2,&      (a&=1,\ b=\epsilon,\ c=0,\ d=1)
\label{3e41}
\end{alignat}
where $\epsilon$ is an infinitesimal parameter. Denoting $\delta x=x'-x$, each infinitesimal transformations are represented by generators $Q_1$, $Q_2$ and $Q_3$ as follows: 
\begin{alignat}{2}
{\rm i)}  \ &\delta x=\epsilon    =[\epsilon Q_1,x],&\quad Q_1&=   \frac{\partial}{\partial x}
\label{3e42}\\
{\rm ii)} \ &\delta x=\epsilon x  =[\epsilon Q_2,x],&      Q_2&=x  \frac{\partial}{\partial x}
\label{3e43}\\
{\rm iii)}\ &\delta x=\epsilon x^2=[\epsilon Q_3,x].&\quad      Q_3&=x^2\frac{\partial}{\partial x}
\label{3e44}
\end{alignat}
Commutation relations $[Q_3, Q_2]=-Q_3$, $[Q_1, Q_2]=Q_1$, $[Q_3, Q_1]=-2 Q_2$ can be modified into the following form
\begin{equation}
[i Q_3, Q_2]=-i Q_3, \quad [i Q_1, Q_2]=i Q_1, \quad [i Q_3, i Q_1]=2 Q_2.
\label{3e45}
\end{equation}
On the other hand, the Lie algebra of SO(3), $[J_a, J_b]= i \epsilon_{a b c} J_c$,
can be rewritten in the form
\begin{equation}
[J_{+}, J_3]=-J_{+}, \quad [J_{-}, J_3]=J_{-}, \quad [J_{+}, J_{-}]=2 J_3,
\nonumber
\end{equation}
with $J_{\pm}=J_1 \pm i J_2$. Then we have the correspondence
$$
J_+ \leftrightarrow i Q_3,\quad
J_- \leftrightarrow i Q_1,\quad
J_3 \leftrightarrow Q_2,
$$
which gives the SO(2,1) Lie algebra structure.

For our purpose to fix $k_{ij}$ which are coefficients of the constant shift of $\wp_{ij}$, and $\ell_i$ which are the coefficients of quadratic invariance, it is sufficient to consider the infinitesimal transformation i)
\begin{equation}
x'_1=x_1+\epsilon, \quad
x'_2=x_2+\epsilon, \quad
y'_1=y_1, \quad
y'_2=y_2 . 
\label{3e46}
\end{equation}
In this case, $a_i$ transform as
\begin{alignat}{4}
a_6'&=a_6,&\quad
a_5'&=a_5- \epsilon a_6,&\quad
a_4'&=a_4-2\epsilon a_5,&\quad
a_3'&=a_3-3\epsilon a_4,
\nonumber\\
a_2'&=a_2-4\epsilon a_3,&
a_1'&=a_1-5\epsilon a_2,&
a_0'&=a_0-6\epsilon a_1.
\label{3e47}
\end{alignat}
Transformation laws of $\wwp_{ij}$ and $\wwp_{ijkl}$ are determined in the following way
\begin{equation}
\widehat{\wp}_{22}'=\widehat{\wp}_{22},                             \quad
\widehat{\wp}_{21}'=\widehat{\wp}_{21}- \epsilon \widehat{\wp}_{22},\quad
\widehat{\wp}_{11}'=\widehat{\wp}_{11}-2\epsilon \widehat{\wp}_{21},
\label{3e48}
\end{equation}
and
\begin{alignat}{3}
\wwp_{2222}'&=\wwp_{2222},&\quad
\wwp_{2221}'&=\wwp_{2221}- \epsilon \wwp_{2222},&\quad
\wwp_{2211}'&=\wwp_{2211}-2\epsilon \wwp_{2221},
\nonumber\\
\wwp_{2111}'&=\wwp_{2111}-3\epsilon \wwp_{2211},&
\wwp_{1111}'&=\wwp_{1111}-4\epsilon \wwp_{2111}.
\label{3e49}
\end{alignat}
We put $a_6=a'_6=0$ after the transformation.

In order to determine $k_{ij}$, we can use the infinitesimal transformation of differential equations. The infinitesimal transformation of Eq.({\ref{3e47}) raise the order of differential equations, that is, we obtain
Eq.(\ref{3e21}), Eq.(\ref{3e20}), Eq.(\ref{3e19}), and Eq.(\ref{3e18})  from Eq.(\ref{3e22}), 
Eq.(\ref{3e21}), Eq.(\ref{3e20}), and Eq.(\ref{3e19}), respectively. Using this method, we reproduce Eq.(\ref{3e17})~\cite{Baker5}. Here, we demonstrate another method to use the fundamental relation, which generate all differential equation by differentiation, of the form
\begin{equation}
\wp_{22} (u_1, u_2) x_1 x_2 +\wp_{21} (u_1, u_2) (x_1+x_2)
+\wp_{11}(u_1, u_2)=\frac{F(x_1,x_2)-2 y_1 y_2}{4(x_1-x_2)^2}.
\label{3e50}
\end{equation}
This relation is trivially satisfied by using
$$
\wp_{22}(u_1,u_2)= \frac{\lambda_5}{4} (x_1+x_2),\quad
\wp_{21}(u_1,u_2)=-\frac{\lambda_5}{4}  x_1 x_2 ,\quad
\wp_{11}(u_1,u_2)= \frac{F(x_1,x_2)-2y_1y_2}{4(x_1-x_2)^2}.
$$
By using the shifted $\wp_{ij}$ functions of Eqs.(\ref{3e9})--(\ref{3e11}), Eq.(\ref{3e50}) becomes in the form
\begin{align}
& (\wwp_{22}(u_1, u_2)-k_{22} a_4) x_1 x_2 
+(\wwp_{21}(u_1, u_2)-k_{21} a_3) (x_1+x_2)
+(\wwp_{11}(u_1, u_2)-k_{11} a_2)
\nonumber 
\\
&=\frac{F(x_1,x_2)-2 y_1 y_2}{4(x_1-x_2)^2} .
\label{3e51}
\end{align}
Coefficients $k_{ij}$ are determined by the invariance of this fundamental relation under the infinitesimal transformation as follows
$$
k_{22}=\frac32,\quad
k_{21}=\frac12,\quad
k_{11}=\frac32,
$$
which reproduces Eq.(\ref{3e17}). This is the necessary condition that the fundamental relation is invariant under the finite transformation.

Next, we determine $\ell_i$ by using Eq.(\ref{3e35}). By imposing the invariance of Eq.(\ref{3e35}) under the infinitesimal transformation, we obtain 
\begin{align}
I \rightarrow I'&=\ell_1 X_1 (X_5-4 \epsilon X_4)+\ell_2 (X_2- \epsilon X_1) (X_4-3 \epsilon X_3)
+\ell_3 (X_3- 2 \epsilon X_2)^2
\nonumber\\
&=I-\epsilon ( (4 \ell_1+\ell_2) X_1 X_4+(3 \ell_2+4 \ell_3) X_2 X_3)=I,
\nonumber
\end{align}
which gives $\ell_1:\ell_2:\ell_3=1:-4:3$. This is the necessary condition that $I$ is invariant under the finite transformation. Thus, we obtain
\begin{equation}
I=X_1 X_5-4 X_2 X_4+3 X_3^2=\text{inv.},
\label{3e52}
\end{equation}
which reproduces Eq.(\ref{3e36}).

%\newpage
%\vspace{10mm}
%%%%%%%%%%%%%%%%%%%%%%%%%%%%%%%%%%%%%%%%%%%%%%%%%%%%%%%%%%%%%%%%%%%
%%%%%%%%%%%%%%%%%%%%%%%%%%%%%%%%%%%%%%%%%%%%%%%%%%%%%%%%%%%%%%%%%%%
\section{
The Lie group structure of genus three hyperelliptic $\wp_{ij}$ functions} 
\setcounter{equation}{0}
%
%%%%%%%%%%%%%
We parametrize the genus three hyperelliptic curve on $\mathbb{R}$ in the form
\begin{align}
y^2&=\sum_{n=0}^8 \lambda_n x^n=\sum_{n=0}^8 {}_8 C_n a_n x^n  \nonumber\\
   &=a_8 x^8+8 a_7 x^7+28 a_6 x^6+56 a_5 x^5+70 a_4 x^4+56 a_3 x^3+28 a_2 x^2+8 a_1 x+a_0, 
\label{4e1}
\end{align}
and put $a_8=0$ in the end. The Jacobi's inversion problem is given by relations
\begin{equation}
\dd u_1 =\sum_{i=1}^3 \frac{\dd x_i}{y_i}, \quad 
\dd u_2 =\sum_{i=1}^3\frac{ x_i \dd x_i}{y_i}, \quad
\dd u_3 =\sum_{i=1}^3\frac{ x_i^2 \dd x_i}{y_i}. 
\label{4e2}
\end{equation}
Then we have 
\begin{eqnarray}
\frac{ \partial x_1}{\partial u_3} = \frac{y_1}{(x_1-x_2)(x_1-x_3)}, \quad
\frac{ \partial x_1}{\partial u_2} =-\frac{(x_2+x_3)y_1}{(x_1-x_2)(x_1-x_3)}, \quad
\frac{ \partial x_1}{\partial u_1} = \frac{(x_2 x_3)y_1}{(x_1-x_2)(x_1-x_3)}, \quad
\label{4e3}
\end{eqnarray}
and relations with $\{x_1, x_2, x_3\}$, $\{y_1, y_2, y_3\}$ cyclic permutations. It is convenient to obtain $\wp_{ij}(i, j=1, 2, 3)$ from $\zeta_i, (i=1, 2, 3)$ functions. The $\zeta_i$ functions are given by~\cite{Baker1}
\begin{align}
\dd(-\zeta_3)=
 &\sum_{i=1}^3 \frac{ \left(2 \lambda_8 x_i^4+\lambda_7 x_i^3\right)\dd x_i}{4y_i}  ,
\label{4e4}\\
\dd(-\zeta_2)=
 &\sum_{i=1}^3 \frac{ \left(4 \lambda_8 x_i^5+3 \lambda_7 x_i^4+2 \lambda_6 x_i^3+\lambda_5 x_i^2\right)\dd x_i}{4y_i}
\nonumber\\
 &-\frac{1}{2}\dd\left(\frac{y_1}{(x_1-x_2)(x_1-x_3)}
            +\frac{y_2}{(x_2-x_1)(x_2-x_3)}
            +\frac{y_3}{(x_3-x_1)(x_3-x_2)}\right)  ,
\label{4e5}\\
\dd(-\zeta_1)=
 &\sum_{i=1}^3 \frac{ \left(6\lambda_8 x_i^6+5\lambda_7 x_i^5+4\lambda_6 x_i^4+3\lambda_5 x_i^3
                           +2\lambda_4 x_i^2+ \lambda_3 x_i \right)\dd x_i}{4y_i}
\nonumber\\
 &-\frac{1}{2}\dd\left(\frac{(x_1-x_2-x_3) y_1}{(x_1-x_2)(x_1-x_3)}
            +\frac{(x_2-x_3-x_1) y_2}{(x_2-x_1)(x_2-x_3)}
            +\frac{(x_3-x_1-x_2) y_3}{(x_3-x_1)(x_3-x_2)}\right)  .
\label{4e6}
\end{align}
For these $\zeta_i$ functions, we can check the integrability conditions $\partial \zeta_i/\partial u_j=\partial \zeta_j/\partial u_i$, $(1\leq i < j \leq 3)$. In order that differential equations become of the polynomial type, we must put $\lambda_8=0$. In this case, we obtain 
\begin{equation}
\dd(-\zeta_3)=\frac{\lambda_7}{4} \sum_{i=1}^3 \frac{ x_i^3 \dd x_i}{y_i} =\sum_{j=1}^3 \wp_{3j} \dd u_j ,
\label{4e7}
\end{equation}
which provides
\begin{align}
&\wp_{33}=\frac{\partial (- \zeta_3)}{\partial u_3}
=\frac{\lambda_7}{4} (x_1+x_2+x_3)  , 
\label{4e8}\\
&\wp_{32}=\frac{\partial (- \zeta_3)}{\partial u_2}
=-\frac{\lambda_7}{4} (x_1 x_2+x_2 x_3+x_3 x_1)  ,
\label{4e9}\\
&\wp_{31}=\frac{\partial (- \zeta_3)}{\partial u_1}=\frac{\lambda_7}{4} (x_1 x_2 x_3)  .
\label{4e10}
\end{align}
For other $\wp_{ij}$ are defined by using $\zeta_i$ functions in the form $\wp_{ij}=\partial_i (- \zeta_j)(=\partial_j (- \zeta_i))$, $\wp_{ij}=\wp_{ji}$, which satisfy integrability conditions $\partial_i \wp_{jk}=\partial_j \wp_{ik}$. While, from the $\sigma$ function, we can define $\wwp_{ij}$ in the form $\wwp_{ij}=-\partial_i \partial_j \log \sigma$. Though $\wp_{ij}$ and $\wwp_{ij}$ satisfy the same integrability conditions, $\partial_i \wp_{jk}=\partial_j \wp_{ik}$ and $\partial_i \wwp_{jk}=\partial_j \wwp_{ik}$, $\wp_{ij}$ and $\wwp_{ij}$ differ in general by a constant. Thus, by the dimensional analysis, we put in the form $\wp_{ij}=\wwp_{ij}-k_{ij} a_{i+j}$. How to determine coefficients of the constant shift} $k_{ij}$ is explained later. The values of them are given by
\begin{align}
\wp_{33}&=\wwp_{33}-3 a_6,  &  \wp_{32}&=\wwp_{32}-2 a_5, & \wp_{31}&=\wwp_{31}-\frac{1}{2}a_4,
\nonumber\\ 
\wp_{22}&=\wwp_{22}-9 a_4,  & \wp_{21}&=\wwp_{21}-2 a_3, & \wp_{11}&=\wwp_{11}-3 a_2.
\label{4e11} 
\end{align}

Next, we consider following differential equations\cite{Baker3,Baker4,Buchstaber1,Hayashi6}
\begin{align}
 1)&\ \wp_{3333} - 6\wp_{33}^2 + 3\lambda_8\wp_{22} - 4\lambda_8\wp_{31} - \lambda_7\wp_{32} - \lambda_6\wp_{33}
                 + \frac{1}{2}\lambda_8\lambda_4 - \frac{1}{8}\lambda_7\lambda_5 = 0,
\label{4e12}\\
 2)&\ \wp_{3332} - 6\wp_{33}\wp_{32} + 2\lambda_8\wp_{21} + \frac{1}{2}\lambda_7\wp_{22}
                 - \frac{3}{2}\lambda_7\wp_{31} - \lambda_6\wp_{32} + \frac{1}{4}\lambda_8\lambda_3 = 0,
\label{4e13}\\
 3)&\ \wp_{3331} - 6\wp_{33}\wp_{31} - \lambda_8\wp_{11} + \frac{1}{2}\lambda_7\wp_{21} - \lambda_6\wp_{31} = 0, 
\label{4e14}\\
 4)&\ \wp_{3322} - 4\wp_{32}^2 - 2\wp_{33}\wp_{22} + 2\lambda_8\wp_{11} + \frac{1}{2}\lambda_7\wp_{21}
                 - \lambda_6\wp_{31} - \frac{1}{2}\lambda_5\wp_{32} + \frac{1}{2}\lambda_8\lambda_2 = 0,
\label{4e15}\\
 5)&\ \wp_{3321} - 4\wp_{32}\wp_{31} - 2\wp_{33}\wp_{21} - \frac{1}{2}\lambda_5\wp_{31} + \frac{1}{4}\lambda_8\lambda_1 = 0, 
\label{4e16}\\
 6)&\ \wp_{3311} - 4\wp_{31}^2 - 2\wp_{33}\wp_{11} - 2\Delta+\frac{1}{2}\lambda_8\lambda_0 = 0,
\label{4e17} \\
 7)&\ \wp_{3222} - 6\wp_{32}\wp_{22} + \frac{3}{2}\lambda_7\wp_{11} - \lambda_5 \wp_{31} - \lambda_4\wp_{32}
                 + \frac{1}{2}\lambda_3\wp_{33} +\frac{1}{2}\lambda_8\lambda_1 + \frac{1}{4}\lambda_7\lambda_2 = 0, 
\label{4e18}\\
 8)&\ \wp_{3221} - 4\wp_{32}\wp_{21} - 2\wp_{31}\wp_{22} + 2\Delta - \lambda_4\wp_{31}
                 + \frac{1}{2}\lambda_8\lambda_0 + \frac{1}{8}\lambda_7\lambda_1 = 0,
\label{4e19}\\
 9)&\ \wp_{3211} - 4\wp_{31}\wp_{21} - 2\wp_{32}\wp_{11} - \frac{1}{2}\lambda_3\wp_{31} + \frac{1}{4}\lambda_7\lambda_0 = 0,
\label{4e20}\\
10)&\ \wp_{3111} - 6\wp_{31}\wp_{11} - \lambda_2\wp_{31} + \frac{1}{2}\lambda_1\wp_{32} - \lambda_0 \wp_{33} = 0,
\label{4e21}\\
11)&\ \wp_{2222} - 6\wp_{22}^2 -1 2\Delta + 3\lambda_6\wp_{11} - \lambda_5\wp_{21} - \lambda_4\wp_{22}
                 - \lambda_3\wp_{32} + 3\lambda_2\wp_{33} + \lambda_8\lambda_0 + \frac{3}{8}\lambda_7\lambda_1
\nonumber\\
   &\qquad       + \frac{1}{2}\lambda_6 \lambda_2 - \frac{1}{8}\lambda_5\lambda_3 = 0, 
\label{4e22}\\
12)&\ \wp_{2221} - 6\wp_{22}\wp_{21} + \frac{1}{2}\lambda_5 \wp_{11} - \lambda_4\wp_{21} - \lambda_3 \wp_{31}
                 +\frac{3}{2}\lambda_1\wp_{33} + \frac{1}{2}\lambda_7\lambda_0 + \frac{1}{4}\lambda_6\lambda_1 = 0,
\label{4e23}\\
13)&\ \wp_{2211} - 4\wp_{21}^2 - 2\wp_{22}\wp_{11} - \frac{1}{2}\lambda_3 \wp_{21} - \lambda_2 \wp_{31} 
                 + \frac{1}{2}\lambda_1\wp_{32} + 2\lambda_0\wp_{33}+\frac{1}{2}\lambda_6 \lambda_0=0,
\label{4e24}\\
14)&\ \wp_{2111} - 6\wp_{21}\wp_{11} - \lambda_2\wp_{21} + \frac{1}{2}\lambda_1\wp_{22} - \frac{3}{2}\lambda_1\wp_{31}
                 + 2\lambda_0\wp_{32} + \frac{1}{4}\lambda_5\lambda_0 = 0,
\label{4e25}\\
15)&\ \wp_{1111} - 6\wp_{11}^2 - \lambda_2\wp_{11} - \lambda_1\wp_{21} + 3\lambda_0\wp_{22} - 4\lambda_0\wp_{31} 
                 + \frac{1}{2}\lambda_4\lambda_0 - \frac{1}{8}\lambda_3\lambda_1 = 0,
\label{4e26}
\end{align}
where $\Delta=\wp_{32}\wp_{21}-\wp_{31}\wp_{22}-\wp_{33}\wp_{11}+\wp_{31}^2$. We cannot express Eq.(\ref{4e17}), Eq.(\ref{4e19}) and Eq.(\ref{4e22}) in the Hirota form, because these equations contain $\Delta$ term. Constant shifted differential equations are given in Appendix B. From Eqs.(\ref{B1})--(\ref{B15}), as we did before, we define the first, second, third and fourth term of each equation as the component of vectors of ${\bf P}$, ${\bf Q}$, ${\bf R}$ and ${\bf S}$ in such a form as 
\begin{equation}
P_n+Q_n+R_n+S_n=0,\,(n=1, 2, \cdots, 15) , 
\label{4e27}
\end{equation}
which provide Eqs.(\ref{B1})--(\ref{B15}).

Next, we consider the generalized dual transformation of the form
\begin{equation}
x_i'=\frac{a x_i -c}{-b x_i +d} , \quad y_i'=\frac{y_i}{(-b x_i+ d)^4}, \quad
\textrm{with} \quad  ad-bc=1, 
\quad(i=1, 2, 3),
\label{4e28}
\end{equation}
in such a way as such transformation makes the hyperelliptic curve on $\mathbb{R}$ to be invariant. Then $a'_n \,(n=0,1, \cdots, 8)$ are systematically determined from the relation
\begin{equation}
\sum_{n=0}^8 {}_8 C_n a_n (b x' +a)^{8-n} (d x'+c)^{n}=\sum_{n=0}^8 {}_8 C_n a'_n x'^n.
\label{4e29}
\end{equation}
By using Eq.(\ref{4e2}),  we obtain 
\begin{align}
\dd u_1' &=\sum_{i=1}^3 \frac{\dd x_i'}{y_i'}=\sum_{i=1}^3 \frac{(-b x_i +d) ^2\dd x_i}{y_i}
=d^2\ \dd u_1 -2 b d\ \dd u_2 +b^2\ \dd u_3,  
\label{4e30}\\
\dd u_2' &=\sum_{i=1}^3\frac{ x_i' \dd x_i'}{y_i'}=\sum_{i=1}^3 \frac{(a x_i -c)(-b x_i +d)\dd x_i}{y_i}
%\nonumber\\&
=-cd\ \dd u_1 +(ad+bc)\ \dd u_2 -ab\ \dd u_3,  
\label{4e31}\\
\dd u_3' &=\sum_{i=1}^3\frac{ {x_i'}^2 \dd x_i'}{y_i'}=\sum_{i=1}^3 \frac{(a x_i -c)^2\dd x_i}{y_i}
=c^2\ \dd u_1 -2 ac\ \dd u_2 +a^2\ \dd u_3   , 
\label{4e32}
\end{align}
which provide
\begin{align}
\frac{\partial}{\partial u'_1}&=a^2 \frac{\partial}{\partial u_1}+a c \frac{\partial}{\partial u_2} 
+c^2 \frac{\partial}{\partial u_3} , 
\label{4e33}\\
\frac{\partial}{\partial u'_2}&=2 a b \frac{\partial}{\partial u_1}+(a d+b c) \frac{\partial}{\partial u_2}
+2 c d \frac{\partial}{\partial u_3}   ,
\label{4e34}\\
\frac{\partial}{\partial u'_3}&=b^2 \frac{\partial}{\partial u_1}+b d \frac{\partial}{\partial u_2}
+d^2 \frac{\partial}{\partial u_3}   .
\label{4e35}
\end{align}

The simple rule to obtain the transformations of $\wwp_{ij}$ is as follows. We consider
\begin{align*}
P'_1&=a^2 P_1+ac P_2+c^2 P_3, \\
P'_2&=2ab P_1+(ad+bc) P_2+2cd P_3,\\
P'_3&=b^2 P_1+bd P_2+d^2 P_3.
\end{align*}
Take, for example, 
\begin{equation} 
{P'_3}^2=d^4 {P_3}^2+2 b d^3 P_3 P_2+2 b^2 d^2 P_3 P_1 +b^2 d^2 P_2^2 +2 b^3 d P_2 P_1 +b^4 P_1^2  , 
\label{4e36}
\end{equation}
and replace $P'_i P'_j  \rightarrow \wwp'_{ij}$, $P_i P_j  \rightarrow \wwp_{ij}$. Then, we obtain
\begin{equation}
\wwp_{33}'=d^4 \wwp_{33}+2 b d^3 \wwp_{32}+2 b^2 d^2\wwp_{31} +b^2 d^2\wwp_{22}
+2 b^3 d \wwp_{21}+b^4 \wwp_{11}  .
\label{4e37}
\end{equation}
The transformation of other $\wwp_{ij}$ are obtained in the same way. This simplified rule is also valid for the transformations of $\wwp'_{ijk\ell}$. They are determined by considering $P'_i P'_j P'_k P'_{\ell}$. Thus, we can systematically obtain the transformations of $\wwp'_{ijk\ell}$. For example, by considering ${P'_3}^4$, we obtain
\begin{align}
\wwp'_{3333}=&d^8 \wwp_{3333}+4 b d^7 \wwp_{3332}+4 b^2 d^6 \wwp_{3331} 
+6 b^2 d^6 \wwp_{3322}+12 b^3 d^5 \wwp_{3321}
\nonumber\\
&+6 b^4 d^4 \wwp_{3311}
+4 b^3 d^5 \wwp_{3222}+12 b^4 d^4 \wwp_{3221}+12 b^5 d^3 \wwp_{3211}+4 b^6 d^2 \wwp_{3111}
\nonumber\\
&+b^4 d^4 \wwp_{2222}+4 b^5 d^3 \wwp_{2221}+6 b^6 d^2 \wwp_{2211}+4 b^7 d \wwp_{2111}+b^8 \wwp_{1111}.
\nonumber
\end{align}
We can systematically obtain the transformation of ${\bf P}$, and we denote this as ${\bold P}'=M {\bold P}$. Then, by using the same matrix $M$, we can prove that ${\bold Q}'=M {\bold Q}$. We determined $k_{ij}$ of the constant shift $\wp_{ij}=\wwp_{ij}-k_{ij} a_{i+j}$ in such a way as ${\bold R}$ transform in the same way as ${\bold R}'=M {\bold R}$, which gives
$$
k_{33}=3,\quad
k_{32}=2,\quad
k_{31}=\frac12,\quad
k_{22}=9,\quad
k_{21}=2,\quad
k_{11}=3.
$$
Thus, we obtain the constant shift of Eq.(\ref{4e11}). After fixing $M$ and $k_{ij}$, we can prove ${\bold S}'=M {\bold S}$. 

We define the total vector ${\bf T}={\bf P}+{\bf Q}+{\bf R}+{\bf S}$, and the whole differential equations transform covariantly in the form ${\bold T}'=M {\bold T}$. 
In the genus three case, some differential equations cannot be expressed in the Hirota form because of the $\widehat{\Delta}$ term. Nevertheless, we find the whole differential equations transform covariantly, since $\widehat{\Delta}$ itself is invariant under the transformation.  

We have shown that differential equations have the Lie group (continuous group) structure; yet an issue is what type of Lie group structure differential equations have. For that purpose, we try to find quadric invariances of some vector ${\bf X}$. We here adopt ${\bf X}={\bf P}$. By the dual transformation, $\wwp_{ijk\ell} \leftrightarrow \wwp_{4-i,4-j,4-k,4-\ell}$, we take the following combination to obtain a quadratic invariance,
\begin{align}
I=&
 \ell_1 \wwp_{3333} \wwp_{1111}
+\ell_2 \wwp_{3332} \wwp_{2111} 
+\ell_3 \wwp_{3331} \wwp_{3111}
+\ell_4 \wwp_{3322} \wwp_{2211} 
+\ell_5 \wwp_{3321} \wwp_{3211} 
+\ell_6 \wwp_{3311}^2  
\nonumber\\
&
+\ell_7 \wwp_{3222} \wwp_{2221}
+\ell_8 \wwp_{3221}^2
+\ell_9 \wwp_{2222}^2
\nonumber\\
=&
 \ell_1 X_1 X_{15}
+\ell_2 X_2 X_{14}
+\ell_3 X_3 X_{10}
+\ell_4 X_4 X_{13}
+\ell_5 X_5 X_{9}
+\ell_6 X_6^2 
\nonumber\\
&+\ell_7 X_7 X_{12}
+\ell_8 X_8^2
+\ell_9 X_{11}^2  , 
\label{4e38}
\end{align}
where we denote $X_n (n=1,2,\cdots,15)$ as the first term of the $n)'$-th equation in Appendix B. By imposing that $I$ is invariant under ${\bold P}'=M {\bold P}$, coefficients $\ell_i$ are determined in the form
\begin{align}
I=&
  2 \wwp_{3333} \wwp_{1111}
 -4 \wwp_{3332} \wwp_{2111} 
 +8 \wwp_{3331} \wwp_{3111}
 +3 \wwp_{3322} \wwp_{2211} 
-12 \wwp_{3321} \wwp_{3211} 
 +6 \wwp_{3311}^2  
\nonumber\\
&
  - \wwp_{3222} \wwp_{2221}
 +3 \wwp_{3221}^2
 +\frac{1}{16} \wwp_{2222} ^2
\nonumber\\
=&
  2 X_1 X_{15}
 -4 X_2 X_{14}
 +8 X_3 X_{10}
 +3 X_4 X_{13}
-12 X_5 X_{9}
 +6 X_6^2 
\nonumber\\
&
  - X_7 X_{12}+3 X_8^2
 +\frac{1}{16} X_{11}^2  .
\label{4e39}
\end{align}
Even if we adopt ${\bf X}$ as any of $\{ {\bf P},  {\bf Q},  {\bf R},  {\bf S} \}$, Eq.(\ref{4e39}) gives invariants. Thus, we define
\begin{alignat}{3}
X_{1\hphantom{0}}&=(Y_1+Y_{10})/\sqrt{2},& \qquad
X_{2\hphantom{0}}&=(Y_{11}+Y_2)/2,       & \qquad
X_{3\hphantom{0}}&=(Y_3+Y_{12})/\sqrt{8},
\notag\\
X_{4\hphantom{0}}&=(Y_4+Y_{13})/\sqrt{3}, &
X_{5\hphantom{0}}&=(Y_{14}+Y_5)/\sqrt{12},& 
X_{6\hphantom{0}}&=Y_{7}/\sqrt{6}, 
\notag\\
X_{7\hphantom{0}}&=Y_{15}+Y_6,    &
X_{8\hphantom{0}}&=Y_{8}/\sqrt{3},& 
X_{9\hphantom{0}}&=(Y_{14}-Y_5)/\sqrt{12},
\notag\\
X_{10}&=(Y_3-Y_{12})/\sqrt{8}, & 
X_{11}&=4 Y_{9},               &
X_{12}&=Y_{15}-Y_6, 
\notag\\
X_{13}&=(Y_4-Y_{13})/\sqrt{3},&
X_{14}&=(Y_{11}-Y_2)/2,       & 
X_{15}&=(Y_1-Y_{10})/\sqrt{2}.
\label{4e40}
\end{alignat}
Then we obtain the invariance of the form
\begin{align}
I=&\left(Y_1^2+Y_2^2+Y_3^2+Y_4^2+Y_5^2 +Y_6^2+Y_7^2+Y_8^2+Y_9^2\right) \nonumber\\
  &-\left(Y_{10}^2+Y_{11}^2+Y_{12}^2+Y_{13}^2+Y_{14}^2+Y_{15}^2\right)=\text{inv.},
\label{4e41}
\end{align}
which implies that we have the SO(9,6) Lie group and/or it's subgroup structure. While, if we denote ${\bf P}_1=(\wwp_{33}, \wwp_{32}, \wwp_{31},\wwp_{22},\wwp_{21}, \wwp_{11})^T$, and express $\widehat{\Delta}$ in the form $\widehat{\Delta}={\bf P}^T_1 D {\bf P}_1$ with 
\begin{equation}
D
=\left( \begin{array}{@{\,}cccccc@{\,}}
 0  & 0   & 0   & 0   & 0   &-1/2 \\
 0  & 0   & 0   & 0   & 1/2 & 0   \\
 0  & 0   & 1   &-1/2 & 0   & 0   \\
 0  & 0   &-1/2 & 0   & 0   & 0   \\
 0  & 1/2 & 0   & 0   & 0   & 0   \\
-1/2& 0   & 0   & 0   & 0   & 0   \\
 \end{array} \right).
\nonumber 
\end{equation}
The eigenvalues of $D$ is given by
$$
 \frac12,\quad
 \frac12,\quad
 \frac{\sqrt{2}+1}{2},\quad
-\frac12,\quad
-\frac12,\quad
-\frac{\sqrt{2}-1}{2}.
$$
Thus, the invariance of $\widehat{\Delta}$ implies the SO(3,3) Lie group structure. There are various invariances such as $16 {\bf X}_6-8 {\bf X}_8+{\bf X}_{11}=\text{inv.}$, which gives $16 \wwp_{3311}-8\wwp_{3221}+\wwp_{2222}=\text{inv.}$ in the ${\bf X}={\bf P}$ case.

%%%%%%%%%%%%%%%%%%%%%%%%%%%%%%%%%%%%%%%%%%%%%%%%%%%%%%%%%
\subsection{The Lie algebraic Approach to the constant shift of $\wp$ and the quadratic invariant}
We can determine $k_{ij}$ for the constant shift of $\wp_{ij}$ by imposing the invariance of differential equations under the infinitesimal transformation. One of the infinitesimal transformations of Eq.({\ref{4e28}) in the form 
\begin{equation}
x'_i=x_i+\epsilon, \quad y'_i=y_i, \quad (i=1,2,3), 
\label{4e42}
\end{equation}
raise the order of differential equations, that is, we obtain Eq.(\ref{4e25}) from Eq.(\ref{4e26}), Eq.(\ref{4e24}) from Eq.(\ref{4e25}), {\it etc}. Using this method, we can determine coefficients $k_{ij}$ for the constant shift and we reproduce Eq.(\ref{4e11})~\cite{Baker5}. Here, we use another method to use the fundamental relation. The fundamental relation, which generates all differential equations, is given by~\cite{Baker1}
\begin{align}
\sum_{i, j=1}^3 \wp_{i j} (u) x_r^{i-1} x_s^{j-1}
=&
\wp_{33}(u) x_r^2 x_s^2+\wp_{32}(u) x_r x_s (x_r+x_s)  +\wp_{31}(u) (x_r^2+x_s^2)
+\wp_{22}(u) {x_r} {x_s}
\nonumber\\
&+\wp_{21}(u) (x_r+x_s) 
+\wp_{11} (u)
\nonumber 
\\
=&\frac{F(x_r,x_s)-2 y_r y_s}{4(x_r-x_s)^2},\quad 
(1\le r < s \le 3),
\label{4e43}
\end{align}
with
\begin{align*}
F(x_r,x_s)=&2 a_8 x_r^4 x_s^4+8 a_7 x_r^3 x_s^3 (x_r+x_s)+56 a_6 x_r^3 x_s^3
+56 a_5 x_r^2 x_s^2 (x_r+x_s)
+140 a_4 x_r^2 x_s^2
\\
&+56 a_3 x_r x_s (x_r+x_s)+56 a_2 x_r x_s+8 a_1 (x_r+x_s)+2 a_0. 
\end{align*}
For example, taking $x_r=x_1$ and $x_s=x_2$, we obtain
\begin{align}
\sum_{i, j=1}^3 \wp_{i j} (u) x_1^{i-1} x_2^{j-1}
=&\wp_{33}(u) x_1^2 x_2^2+\wp_{32}(u) x_1 x_2 (x_1+x_2)  +\wp_{31}(u) (x_1^2+ x_2^2)
\nonumber\\
&+\wp_{22}(u) {x_1} {x_2}+\wp_{21}(u) (x_1+x_2)  +\wp_{11}(u) 
\nonumber 
\\
=&\frac{F(x_1,x_2)-2 y_1 y_2}{4(x_1-x_2)^2}. 
\label{4e44}
\end{align}
Though Eq.(\ref{4e43}) is not symmetric in the index of $x_1, x_2, x_3$ and $y_1, y_2, y_3$, this fundamental relation is satisfied for $\wp_{ij}(u)=\partial_i (-\zeta_j(u))(=\partial_j (-\zeta_i(u))$. By using shifted $\wwp_{ij}$ functions, we obtain the following from Eq.(\ref{4e44}), 
\begin{align}
&(\wwp_{33}(u)-k_{33} a_6)x_1^2 x_2^2
+(\wwp_{32}(u) -k_{32} a_5) x_1 x_2 (x_1 + x_2)
+(\wwp_{31}(u) -k_{31} a_4)(x_1^2 + x_2^2)
\nonumber\\
&
+(\wwp_{22}(u)-k_{22} a_4) x_1x_2
+(\wwp_{21}(u)-k_{21} a_3)(x_1 + x_2)
+(\wwp_{11}(u)-k_{11} a_2)
\nonumber\\
& 
=\frac{F(x_1,x_2)-2 y_1 y_2}{4(x_1-x_2)^2}. 
\label{4e45}
\end{align}

Under the infinitesimal transformation of Eq.\eqref{4e42}, we have
\begin{alignat}{5}
a_8'&=a_8,&\quad
a_7'&=a_7- \epsilon a_8,&\quad
a_6'&=a_6-2\epsilon a_7,&\quad
a_5'&=a_5-3\epsilon a_6,&\quad
a_4'&=a_4-4\epsilon a_5,
\nonumber\\
a_3'&=a_3-5\epsilon a_4,&
a_2'&=a_2-6\epsilon a_3,&
a_1'&=a_1-7\epsilon a_2,&
a_0'&=a_0-8\epsilon a_1,
\label{4e46}
\end{alignat}
and 
\begin{alignat}{3}
\wwp'_{33}&=\wwp_{33},& \quad
\wwp'_{32}&=\wwp_{32}-2\epsilon \wwp_{33},& \quad
\wwp'_{31}&=\wwp_{31}- \epsilon \wwp_{32},
\nonumber\\
\wwp'_{22}&=\wwp_{22}-4\epsilon \wwp_{32},&
\wwp'_{21}&=\wwp_{21}- \epsilon (\wwp_{22}+2\wwp_{31}),&
\wwp'_{11}&=\wwp_{11}-2\epsilon \wwp_{21}.
\label{4e47}
\end{alignat}
We put $a_8=a'_8=0$ after the transformation. By imposing the infinitesimal invariance of the fundamental relation Eq.(\ref{4e45}), we reproduce Eq.(\ref{4e11}) as the necessary condition of the invariance under the finite transformation.

Next, we construct the quadratic invariance. For that purpose, we consider the combination of  Eq.(\ref{4e38}). By imposing the infinitesimal transformation  
\begin{alignat}{2}
  1)\ \wwp_{3333}'&=\wwp_{3333},& \quad
  2)\ \wwp_{3332}'&=\wwp_{3332}-2\epsilon \wwp_{3333},
\nonumber\\
  3)\ \wwp_{3331}'&=\wwp_{3331}- \epsilon \wwp_{3332}, 
& 4)\ \wwp_{3322}'&=\wwp_{3322}-4\epsilon \wwp_{3332},
\nonumber\\
  5)\ \wwp_{3321}'&=\wwp_{3321}- \epsilon(\wwp_{3322}+2\wwp_{3331}), 
& 6)\ \wwp_{3311}'&=\wwp_{3311}-2\epsilon \wwp_{3321}, 
 \nonumber\\
  7)\ \wwp_{3222}'&=\wwp_{3222}-6\epsilon \wwp_{3322}, 
& 8)\ \wwp_{3221}'&=\wwp_{3221}- \epsilon(\wwp_{3222}+4\wwp_{3321}),
\nonumber\\
  9)\ \wwp_{3211}'&=\wwp_{3211}-2\epsilon(\wwp_{3221}+ \wwp_{3311}),  
&10)\ \wwp_{3111}'&=\wwp_{3111}-3\epsilon \wwp_{3211},
\nonumber\\
 11)\ \wwp_{2222}'&=\wwp_{2222}-8\epsilon \wwp_{3222}, 
&12\ )\wwp_{2221}'&=\wwp_{2221}- \epsilon(\wwp_{2222}+6\wwp_{3221}), 
\nonumber\\
 13)\ \wwp_{2211}'&=\wwp_{2211}-2\epsilon(\wwp_{2221}+2\wwp_{3211}), 
&14)\ \wwp_{2111}'&=\wwp_{2111}-\epsilon(3\wwp_{2211}+2\wwp_{3111}), 
\nonumber\\
 15)\ \wwp_{1111}'&=\wwp_{1111}-4\epsilon \wwp_{2111}, 
\label{4e48}
\end{alignat}
we obtain the quadratic invariance of Eq.(\ref{4e39}). This is the necessary condition that Eq.(\ref{4e39}) is invariant under the finite transformation.

%\newpage
%\vspace{10mm}
%%%%%%%%%%%%%%%%%%%%%%%%%%%%%%%%%%%%%%%%%%%%%%%%%%%%%%%%%%%%%%%%%% 
%%%%%%%%%%%%%%%%%%%%%%%%%%%%%%%%%%%%%%%%%%%%%%%%%%%%%%%%%%%%%%%%%% 
\section{Summary and Discussions} 
\setcounter{equation}{0}

In the previous study, by directly using differential equations of genus two hyperelliptic $\wp$ functions, we  demonstrated that the half-period addition formula for genus two hyperelliptic $\wp$ functions provides the order two Sp(4,$\R$) Lie group structure. In this study, we have considered the generalized dual transformation for elliptic/hyperelliptic $\wp$ functions up to genus three. By the constant shift of $\wp_{ij}$ functions, we have deduced that differential equations transform covariantly under such transformation. Thus, we conclude that elliptic/hyperelliptic $\wp$ functions have Lie group structure.

For the genus one case, from the transformation formula, we have deduced that the Weierstrass $\wp$ function has the SO(2,1) $\cong$ Sp(2,$R$)/$\Z_2$ Lie group structure. We have also shown that hyperelliptic $\wp$ functions posses the SO(3,2) $\cong$ Sp(4,$\R$)/$\Z_2$ Lie group structure for genus two case. Further, we have shown that hyperelliptic $\wp$ functions posses the SO(9,6) Lie group and/or it's subgroup structure , which is expected to be Sp(6,$\R$) Lie group structure, for genus three case.

The reason for our expectation is as follows. Let us consider a special case where the off-diagonal moduli of genus three hyperelliptic theta functions vanish in the form $\tau_{i j}=0$ $(1\le i < j \le 3)$. Then, they become the product of genus one elliptic theta functions in the form
$$
 \vartheta(u_1,u_2,u_3;\{\tau_{ij}\})
=\vartheta(u_1; \tau_{11})\vartheta(u_2;\tau_{22})\vartheta(u_3;\tau_{33}),
$$ 
which gives
$$
 \log \sigma(u_1, u_2, u_3;\{\tau_{ij}\})
=\log \sigma(u_1;\tau_{11})
+\log \sigma(u_2;\tau_{22})
+\log \sigma(u_3;\tau_{33}).
$$
For each genus one elliptic theta function, we obtain the SO(2,1) $\cong$ Sp(2,$\R$)/$\Z_2$ Lie group structure. Thus, in the genus three case, the Lie group is expected to contain SO(2,1) $\otimes$ SO(2,1) $\otimes$ SO(2,1) as a subgroup. Hence, we expect to obtain the Sp(6,$\R$) Lie group structure because Sp(6,$\R$) contains Sp(2,$\R$) $\otimes$ Sp(2,$\R$) $\otimes$ Sp(2,$\R$) as a subgroup. In the genus three case, by using other method, we expect that the Lie group structure reduces from SO(9,6) to Sp(6,$\R$).

In the genus three case, though some differential equations contain $\widehat{\Delta}$, which is not expressed by the Hirota form, the whole differential equations have the Lie group structure. Thus, there may be the Lie group structure for any genus hyperelliptic $\wp$ functions.

%%%%%%%%%%%%%%%%%%%%%%%%%%%%%%%%%%%%%%%%%%%%%%%%%%%%%%%%%%%%%%
%%%%%%%%%%%%%%%%%%%%%%%%%%%%%%%%%%%%%%%%%%%%%%%%%%%%%%%%%%%%%%
%%%%%%%%%%%%%%%%%%%%%%%%%%%%%%%%%%%%%%%%%%%%%%%%%%%%%%%%%%%%%%
\begin{appendices}
\setcounter{equation}{0}
\section{Genus two: the transformation of the coefficients $a_n\,(n=0,1, \cdots, 6)$ in the 
hyperelliptic curve} 
\begin{align}
 1)~a'_6=&   a_6     d^6
         + 6 a_5 b   d^5
         +15 a_4 b^2 d^4
         +20 a_3 b^3 d^3
         +15 a_2 b^4 d^2
         + 6 a_1 b^5 d
         +   a_0 b^6,
\label{A1}\\
 2)~a'_5=&  a_6               cd^5
         +  a_5( ad + 5bc )    d^4
         + 5a_4( ad + 2bc ) b  d^3
         +10a_3( ad +  bc ) b^2d^2 \nonumber\\ 
        &+ 5a_2(2ad +  bc ) b^3d
         +  a_1(5ad +  bc ) b^4
         +  a_0            ab^5
\label{A2}\\
 3)~a'_4=&  a_6                               c^2d^4
         + 2a_5( ad + 2bc )                   c  d^3
         +  a_4( a^2d^2 + 8abcd + 6b^2c^2 )      d^2
         + 4a_3( a^2d^2 + 3abcd +  b^2c^2 )   b  d   \nonumber\\
        &+  a_2(6a^2d^2 + 8abcd +  b^2c^2 )   b^2 
         + 2a_1(2ad + bc )                 a  b^3
         +  a_0                            a^2b^4
\label{A3}\\
 4)~a'_3=&  a_6                            c^3d^3
         + 3a_5( ad + bc ) c^2d^2
         + 3a_4( a^2d^2 + 3abcd + b^2c^2 )c  d
 \nonumber\\
        &+  a_3 (a^3d^3 + 9a^2bcd^2 + 9ab^2c^2d + b^3c^3 )
         + 3a_2( a^2d^2 + 3abcd + b^2c^2 ) a  b
 \nonumber\\
        &+ 3a_1( ad + bc )                 a^2b^2
         +  a_0                            a^3b^3
\label{A4}\\
 5)~a'_2=&  a_6                               c^4d^2
         + 2a_5(2ad + bc )                    c^3d
         +  a_4(6a^2d^2 + 8abcd +  b^2c^2)    c^2
         + 4a_3( a^2d^2 + 3abcd +  b^2c^2) a  c \nonumber\\
        &+  a_2( a^2d^2 + 8abcd + 6b^2c^2) a^2
         + 2a_1( ad + 2bc)                 a^3b
         +  a_0                            a^4b^2
\label{A5}\\
 6)~a'_1=&  a_6                c^5d
         +  a_5(5ad +  bc )    c^4
         + 5a_4(2ad +  bc ) a  c^3
         +10a_3( ad +  bc ) a^2c^2  \nonumber\\
        &+ 5a_2( ad + 2bc ) a^3c
         +  a_1( ad + 5bc ) a^4
         +  a_0             a^5b
\label{A6}\\
 7)~a'_0=&  a_6 c^6+6 a_5 a c^5
         +15a_4a^2c^4
         +20a_3a^3c^3
         +15a_2a^4c^2
         + 6a_1a^5c
         +  a_0a^6.
\label{A7}
\end{align}
We put $a_6=0$ and $a'_6=0$ after the transformation. For example, we can realize $a'_6=0$ by taking the standard form of the hyperelliptic curve with $a_0=-(6 a_5 d^5+15 a_4 b d^4+20 a_3 b^2 d^3 +15 a_2 b^3 d^2+6 a_1 b^4 d)/b^5 $.

%%%%%%%%%%%%%%%%%%%%%%%%%%%%%%%%%%%%%%%%%%%%%%%%%%%%%%%%%%%%%%%%%%%
%%%%%%%%%%%%%%%%%%%%%%%%%%%%%%%%%%%%%%%%%%%%%%%%%%%%%%%%%%%%%%%%%%%
\setcounter{equation}{0}
\section{Genus three:  shifted differential equations of $\wp_{ijk\ell}$} 
\begin{align}
 1)'&\ \wwp_{3333}-6\wwp_{33}^2+(3 a_8 \wwp_{22}-4 a_8 \wwp_{31}-8 a_7 \wwp_{32}+8 a_6\wwp_{33})
\nonumber\\
&\hskip 7mm +(10 a_8 a_4 -40 a_7 a_5+30 a_6^2)=0 ,
\label{B1}\\
 2)'&\ \wwp_{3332}-6 \wwp_{33} \wwp_{32}+(2 a_8 \wwp_{21}+4 a_7 \wwp_{22}-12 a_7 \wwp_{31} 
-10 a_6\wp_{32}+12 a_5 \wwp_{33})
\nonumber\\
&\hskip 7mm +(10 a_8 a_3 -30 a_7 a_4 +20 a_6 a_5)=0 ,
\label{B2}\\
 3)'&\ \wwp_{3331}-6\wwp_{33}\wwp_{31}+(-a_8 \wwp_{11} +4a_7 \wwp_{21} - 10 a_6\wwp_{31}+3 a_4 \wwp_{33})
\nonumber\\
&\hskip 7mm +(3 a_8 a_2-8 a_7 a_3 +5 a_6 a_4)=0, 
\label{B3}\\
 4)'&\ \wwp_{3322}-(4 \wwp_{32}^2+2\wwp_{33} \wwp_{22})+(2 a_8 \wwp_{11}
+4 a_7\wwp_{21}+6 a_6 \wwp_{22}-28 a_6 \wwp_{31}-12 a_5\wwp_{32} +18 a_4 \wwp_{33})
\nonumber\\
&\hskip 7mm +(8 a_8 a_2 -8 a_7 a_3 -40 a_6 a_4 +40 a_5^2)=0,
\label{B4}\\
 5)'&\ \wwp_{3321}-(4 \wwp_{32}\wwp_{31}+2\wwp_{33} \wwp_{21})+(6 a_6 \wwp_{21}-20 a_5\wp_{31} 
+2 a_4\wwp_{32}+4 a_3\wwp_{33})
\nonumber\\
&\hskip 7mm +(2 a_8 a_1-12 a_6 a_3 +10 a_5 a_4)=0 , 
\label{B5}\\
 6)'&\ \wwp_{3311}-\left(4 \wwp_{31}^2 +2 \wwp_{33}\wwp_{11}+2 \widehat{\Delta}\right)
+(4 a_5 \wwp_{21}-a_4 \wwp_{22}-12 a_4 \wwp_{31}+4 a_3 \wwp_{32})
\nonumber\\
&\hskip 7mm +\left(\frac{1}{2} a_8 a_0 -8 a_5 a_3+\frac{15}{2} a_4^2\right)=0,
\label{B6}\\
 7)'&\ \wwp_{3222}-6\wwp_{32} \wwp_{22}+(12 a_7\wwp_{11}+12 a_5\wwp_{22}  -56 a_5 \wwp_{31}
-16 a_4\wwp_{32}+28 a_3\wwp_{33})
\nonumber\\
&\hskip 7mm +(4 a_8 a_1+20 a_7 a_2-84 a_6 a_3 +60 a_5 a_4)=0 , 
\label{B7}\\
 8)'&\ \wwp_{3221}-\left(4\wwp_{32}\wwp_{21}+2\wwp_{31} \wwp_{22}-2\widehat{\Delta}\right)
\nonumber\\
&\hskip 7mm  +(6 a_6 \wwp_{11}  +4 a_5 \wwp_{21}+2 a_4 \wwp_{22} -36 a_4\wwp_{31}  +4 a_3 \wwp_{32}+6 a_2 \wwp_{33})
\nonumber\\
&\hskip 7mm +\left(\frac{1}{2} a_8 a_0+8 a_7 a_1-18 a_6 a_2-8 a_5 a_3+\frac{35}{2} a_4^2\right)=0,
\label{B8}\\
 9)'&\ \wwp_{3211}-(4 \wwp_{31}\wwp_{21}+2\wwp_{32} \wwp_{11})+(4 a_5 \wwp_{11}+2 a_4 \wwp_{21}
-20 a_3\wwp_{31}+6 a_2 \wwp_{32})
\nonumber\\  
&\hskip 7mm +(2 a_7 a_0-12 a_5 a_2 +10 a_4 a_3)=0 ,
\label{B9}\\
10)'&\ \wwp_{3111}-6\wwp_{31}\wwp_{11}+(3 a_4 \wwp_{11}-10 a_2 \wwp_{31}+4 a_1 \wwp_{32}
-a_0 \wwp_{33})
\nonumber\\
&\hskip 7mm +(3 a_6 a_0 -8 a_5 a_1+5 a_4 a_2)=0 ,
\label{B10}\\
11)'&\ \wwp_{2222}-\left(6\wwp_{22}^2+12\widehat{\Delta}\right)+(48 a_6 \wwp_{11}-32 a_5 \wwp_{21}
+32 a_4\wwp_{22}-96 a_4\wwp_{31}-32 a_3\wwp_{32}+48 a_2\wwp_{33})
\nonumber\\
&\hskip 7mm +(a_8 a_0+24 a_7 a_1-4 a_6 a_2-216 a_5 a_3+195 a_4^2)=0, 
\label{B11}\\
12)'&\ \wwp_{2221}-6\wwp_{22} \wwp_{21}+(28 a_5 \wwp_{11}-16 a_4\wwp_{21} +12 a_3\wwp_{22}
-56 a_3 \wwp_{31} +12 a_1\wwp_{33} )
\nonumber\\
&\hskip 7mm +(4 a_7 a_0+20 a_6 a_1-84 a_5 a_2+60 a_4 a_3)=0 ,
\label{B12}\\
13)'&\ \wwp_{2211}-(4 \wwp_{21}^2+2\wwp_{22}\wwp_{11})+(18 a_4\wwp_{11} -12 a_3\wwp_{21}
+6 a_2\wwp_{22}-28 a_2 \wwp_{31} +4 a_1 \wwp_{32}+2 a_0 \wwp_{33})
\nonumber\\
&\hskip 7mm +(8 a_6 a_0-8 a_5 a_1-40 a_4 a_2+40 a_3^2)=0 ,
\label{B13}\\
14)'&\ \wwp_{2111}-6\wwp_{21}\wwp_{11}+(12 a_3 \wwp_{11} -10 a_2\wwp_{21}+4 a_1\wwp_{22}-12 a_1 \wwp_{31}
+2 a_0\wwp_{32})
\nonumber\\
&\hskip 7mm +(10 a_5 a_0 -30 a_4 a_1 +20 a_3 a_2)=0 ,
\label{B14}\\
15)'&\ \wwp_{1111}-6\wwp_{11}^2+(8 a_2 \wwp_{11}-8 a_1\wwp_{21} +3 a_0 \wwp_{22} -4 a_0 \wwp_{31})
\nonumber\\
&\hskip 7mm +(10 a_4 a_0-40 a_3 a_1 +30 a_2^2)=0 ,
\label{B15}
\end{align}
where $\widehat{\Delta}=\wwp_{32}\wwp_{21}-\wwp_{31}\wwp_{22}-\wwp_{33}\wwp_{11}+\wwp_{31}^2$.

%%%%%%%%%%%%%%%%%%%%%%%%%%%%%%%%%%%%%%%%%%%%%%%%%%%%%%%%%%%%%%%%%
%%%%%%%%%%%%%%%%%%%%%%%%%%%%%%%%%%%%%%%%%%%%%%%%%%%%%%%%%%%%%%%%%
\end{appendices}

%\newpage
%%%%%%%%%%%%%%%%%%%%%%%%%%%%%%%%%%%%%%%%%%%%%%%%%%%%%%%%%%%  

\end{document}